\documentclass[aps,prd,nofootinbib]{revtex4}
\usepackage[colorlinks=true, pdfstartview=FitV, linkcolor=blue, citecolor=red, urlcolor=magenta]{hyperref}
\usepackage[utf8]{inputenc}
\usepackage[english]{babel}
\usepackage[T1]{fontenc}

\usepackage{calligra}
\usepackage{dutchcal} 

\usepackage{amsmath}
\usepackage{amsfonts}
\usepackage{amssymb}
\usepackage{graphicx}
\usepackage{lmodern}
\usepackage{hyperref}
\hypersetup{colorlinks, 
citecolor=blue,
linkcolor=red,
urlcolor=black}
\usepackage[normalem]{ulem}
\usepackage{booktabs}
\usepackage{lmodern}

\newcommand{\be}{\begin{equation}}
\newcommand{\ee}{\end{equation}}
\newcommand{\bea}{\begin{eqnarray}}
\newcommand{\eea}{\end{eqnarray}}

\newcommand{\ben}{\begin{eqnarray}}
\newcommand{\een}{\end{eqnarray}}

\begin{document}

\title{Vacuum energy, temperature corrections and heat kernel coefficients in $(D+1)$-dimensional spacetimes with nontrivial topology}


\author{$^{1}$Herondy Mota}
\email{hmota@fisica.ufpb.br}

\affiliation{$^{1}$Departamento de F\' isica, Universidade Federal da Para\' iba,\\  Caixa Postal 5008, Jo\~ ao Pessoa, Para\' iba, Brazil.}


\begin{abstract}
In this work we make use of the generalized zeta function technique to investigate the vacuum energy, temperature corrections and heat kernel coefficients associated with a scalar field under a quasiperiodic condition in a $(D+1)$-dimensional conical spacetime. In this scenario we find that the renormalized vacuum energy, as well as the temperature corrections, are both zero. The nonzero heat kernel coefficients are the ones related to the usual Euclidean divergence, and also to the nontrivial aspects of the quaisperiodically identified conical spacetime topology. An interesting result that arises in this configuration is that for some values of the quasiperiodic parameter, the heat kernel coefficient associated with the nontrivial topology vanishes. In addition, we also consider the scalar field in a $(D+1)$-dimensional spacetime formed by the combination of a conical and screw dislocation topological defects. In this case, we obtain a nonzero renormalized vacuum energy density and its corresponding temperature corrections. Again, the nonzero heat kernel coefficients found are the ones related to the Euclidean and nontrivial topology divergences. For $D=3$ we explicitly show, in the massless scalar field case, the limits of low and high temperatures for the free energy. In the latter, we show that the free energy presents a classical contribution. 
\end{abstract}
 \maketitle


\section{Introduction}
\label{intro}

Casimir effect is a phenomenon that arises in the realm of Quantum Field Theory as a consequence of boundary conditions imposed on quantum fields, namely, scalar, electromagnetic and spinor fields. In this sense, the quantum modes of a given field in its vacuum state are altered in such a way that a detectable nonzero Casimir force is produced, at least in the electromagnetic case. In fact, an attractive force was predicted in 1948 by Casimir who considered a configuration constituted by an electromagnetic field whose modes are confined between two identical and large perfect parallel conducting plates \cite{Casimir1948dh}. Although not with a great precision, the first attempt to detect this effect was in 1958 by Sparnaay \cite{Sparnaay1958} and confirmed, after several decades, by others \cite{Lamoreaux:1996wh, Lamoreaux1996wh, MohideenRoy1998iz, Bressi:2002fr, PhysRevA.81.052115, PhysRevA.78.020101}. 

In the past years, Casimir like effects arising due to the geometrical and topological aspects of curved spacetimes have been investigated. In these scenarios, the quantum modes in the vacuum state are modified and a nonzero vacuum energy is produced \cite{bordag2009advances, BordagMohideenMostepanenko}.  

An interesting curved spacetime is the one with a conical topology as, for instance, the cosmic string spacetime. The existence of cosmic strings can produce gravitational, astrophysical and cosmological signatures and is predicted in some extensions of the Standard Model of particle physics \cite{VS, hindmarsh}, and in the context of string theory \cite{Copeland:2011dx, Hindmarsh:2011qj}. Cosmic string is also a linear topological defect that is supposed to be formed due to phase transitions in the early Universe. An idealized and static one is characterized by a spacetime with conical topology that has associated to it a planar angle deficit proportional to its linear energy density, $\mu_0$, that is, $\Delta\phi \simeq8\pi G\mu_0$, where $G$ is the Newton's gravitational constant. In condensed matter systems, the cosmic string counterpart is a disclination \cite{Katanaev:1992kh, Puntigam:1996vy}.  In Refs. \cite{BezerradeMello2011nv, MotaDispiration, klecioQuasiPNanotubes, Braganca:2019mvj, Braganca:2014qma, Braganca:2020jci, deFarias:2021qdg, deFarias:2022rju} quantum effects induced by linear topological defects, such as the ones just described, on the vacuum expectation values of the energy-momentum tensor, induced current density and induced density fluctuation of a classical liquid were considered. In Ref. \cite{Kay:1990cr}, the authors pointed out that the Laplace operator is not self-adjoint in a conical spacetime and investigated possible extensions for such operator.

In the context of theories of solid and crystal continuum media there also exists an interesting linear topological defects known as screw dislocation \cite{Puntigam:1996vy}. The combination of a screw dislocation and a cosmic string (disclination) is also a topological defect that has a spacelike helical structure, with chiral properties \cite{Letelier:1995ze, Galtsov:1993ne, tod1994conical}. In this case, there is not only a delta function singularity in the scalar curvature, associated with the conical defect, but also in the torsion. The latter is a characteristic of a screw dislocation and the combination of it with a conical defect, also known as cosmic dispiration \cite{DeLorenci:2002jv, DeLorenci:2003wv}, is formally constructed in the framework of the Einstein-Cartan theory of gravity \cite{Letelier:1995ze}. 

In dealing with the technical calculation involved in the investigation of Casimir effect it is necessary to adopt regularization and renormalization methods to treat the infinity contributions to the vacuum energy arising in this context.  A very elegant and powerful method is the one constructed by means of the generalized zeta function approach \cite{Hawking1977, DowkerCritchley1975tf, aleixo2021thermal}.  The zeta function in this case is defined through the eigenvalues associated with differential operators such as the Laplace-Beltrami one \cite{Elizalde1994book}. The construction of this approach is by making use of the path integral formalism since, by doing so, it is possible to make a connection with thermodynamics, allowing us to calculate the partition function and, as a consequence, also pave the way to obtain temperature corrections to the vacuum energy \cite{Hawking1977, DowkerCritchley1975tf, aleixo2021thermal}. In this framework, an expansion for the heat kernel is generally adopted in order to study the possible divergent structures associated with nonzero heat kernel coefficients \cite{bordag2009advances, Bordag:1998rf, BordagMohideenMostepanenko, Elizalde1994book}. This approach has been adopted since the famous paper of Kac \cite{Kac:1966xd}.

Our objective in the present work is, by using the generalized zeta function method described above, to study the scalar vacuum energy, temperature corrections and nonzero heat kernel coefficients induced by two types of spacetimes with nontrivial topology, namely, a quasiperiodically identified conical spacetime, and also the spacetime describing a cosmic dispiration. Concerning the former, a massive scalar field whose modes propagate in a conical spacetime is subject to obey a quasiperiodic condition in the azimuthal direction. Such a condition generalizes the widely known periodic and antiperiodic conditions and has been considered in different scenarios \cite{QuasiPNanotubes, klecioQuasiPNanotubes, deFarias:2021qdg, deFarias:2022rju, Junior:2023feu, Ferreira:2023uxs, Ferreira:2022eno, Porfirio:2019gdy, Mota:2016eoi}. Hence, we generalize the results obtained in Ref. \cite{cognola:1993qg}, where the authors studied the vacuum energy, temperature corrections and heat kernel coefficients in the periodic case (see also Ref. \cite{Fursaev:1993qk}). In contrast with Ref. \cite{cognola:1993qg}, in our investigation, we take into consideration the renormalization scheme discussed in Refs. \cite{bordag2009advances, Bordag:1998rf}, where the additional requirement that the renormalized vacuum energy must vanish for large masses is necessary to be adopted. All this is also investigated by considering a cosmic dispiration spacetime which, to the best of our knowledge, is studied for the first time here. Both spacetimes with nontrivial topology are considered in $(D+1)$ dimensions. 

The novelty of the system configuration described above, in the case of the quasiperiodically identified conical spacetime, resides in the generalized expressions for the two-point heat kernel function and generalized zeta function where both of them depend on the quasiperiodic parameter. Also, the nonzero heat kernel coefficients found in this context are the usual coefficient that indicates the presence of the Minkowski spacetime divergence contribution, as well as the heat kernel coefficient that indicates a divergence associated with the conical topology of the spacetime. We show that the latter also depends on the quasiperiodic parameter, thus, generalizing the results reported in Ref. \cite{cognola:1993qg, Fursaev:1993qk}. Upon adopting a consistent renormalization procedure, we also show that after subtracting the divergent contributions characterized by the nonzero heat kernel coefficients, the vacuum energy and its temperature corrections vanish. 

Regarding the novelty arising in the context of the cosmic dispiration spacetime, we obtain generalized expressions for the two-point heat kernel function, generalized zeta function, vacuum energy and temperature corrections where all of them depend on the parameters that characterize the topology of the curved background. Again, the nonzero heat kernel coefficients obtained are the ones related to the Minkowski spacetime divergent contribution, as well as the heat kernel coefficient that indicates a divergence associated with the cosmic dispiration spacetime conical topology. We show that after subtracting the divergent contributions characterized by the nonzero heat kernel coefficients, the vacuum energy and its temperature corrections are finite and nonzero. We analyze the limits of high and low temperatures to validate the consistence of our results.

This work is organized as follows. In Sec.\ref{sec2} we overview the necessary elements, in $(D+1)$ dimensions, involved in the generalized zeta function method such as the heat kernel and free energy definitions. In Sec.\ref{sec3}, the vacuum energy, temperature corrections and heat kernel coefficients are obtained in a quaisperiodically identified $(D+1)$ conical spacetime. In Sec.\ref{sec4}, by doing the same analysis, we consider the $(D+1)$ cosmic dispiration spacetime. Finally, in Sec.\ref{sec5} we present our conclusions. In this paper we use natural units $\hbar=c=1$.

%
\section{Generalized zeta function, heat kernel and free energy}
\label{sec2}
%
In this section we shall consider the general aspects involving the use of the zeta function method for temperature corrections to the vacuum energy of a massive scalar field. The zeta function provides a powerful and elegant tool to calculate the vacuum energy at zero temperature and also to find the corresponding nonzero temperature corrections by adopting the Euclidean formalism, with a compact imaginary time $\tau$. The generalized zeta function in this framework is defined as \cite{Hawking1977, Elizalde1994book, Kirsten:2010zp, Nesterenko:2004gzu, aleixo2021thermal}
\begin{equation}
\zeta_{N}(s)=\sum_{\sigma}\lambda^{-s}_{\sigma},
\label{zeta}
\end{equation}
where $\lambda_{\sigma}$ stands for the eigenvalues of a $N$-dimensional Laplace-Beltrami operator $\hat{A}_{N}$ and may not be always discrete, despite the summation symbol representation above. Moreover, the definition in \eqref{zeta} converges for Re($s$) $>$ $N/2$ and can also be analytically extended for Re($s$) $<$ $N/2$, with a pole at $s=N/2$ \cite{Kirsten:2010zp, Nesterenko:2004gzu}. For static spacetimes, the Laplace-Beltrami operator $\hat{A}_N$ can be expressed as 
\begin{eqnarray}
\hat{A}_{N} &=& -\frac{\partial^2}{\partial \tau^2 } - \nabla^2_D + m^2\nonumber\\
&=& - \frac{\partial^2}{\partial \tau^2 } + \hat{A}_{D},
\label{laplace_Beltrame}
\end{eqnarray}
where $N=D+1$, $\nabla^2_D$ is the Laplace operator in $D$ spatial dimensions and $m$ is the mass of the scalar field, which is the one to be considered here. Note that we are making use of the Euclidean representation of the above operator, obtained by performing a Wick rotation, $t\rightarrow -i\tau$, in the ordinary time parameter $t$ present in the Minkowski spacetime \cite{Hawking1977, Elizalde1994book, aleixo2021thermal}.

For a scalar field, $\Phi_{\sigma}(\tau, {\bf r})$, whose modes propagate in a $(D+1)$ Euclidean static spacetime the temperature correction is introduced by imposing a periodicity condition in the imaginary time coordinate, i.e.,
\begin{equation}\begin{split}\label{periodicity}
\Phi_{\sigma}(\tau, {\bf r}) = \Phi_{\sigma}(\tau + \beta, {\bf r}),
\end{split}\end{equation}
where $\beta=\frac{1}{k_BT}$, with $k_B$ being the Boltzmann constant and $T$ the temperature. This means that, by acting the operator in Eq. \eqref{laplace_Beltrame} on the scalar field obeying the periodicity condition above, we can obtain a solution of the form
\begin{equation}\begin{split}\label{sol_time}
\Phi_{\sigma}(\tau, {\bf r}) = e^{-i\omega_n\tau}\varphi_j({\bf r}),\qquad\quad \omega_n=\frac{2\pi n}{\beta},
\end{split}\end{equation}
where the quantum numbers are now represented by $\sigma = (n, j)$ and $n=0, \pm 1, \pm 2,...$\;. This provides a set of eigenvalues as follows
\begin{equation}\begin{split}\label{eigen1}
\lambda_{\sigma} = \left(\frac{2\pi n}{\beta}\right)^2 + \Omega_j,
\end{split}\end{equation}
with $\Omega_j$ being the set of eigenvalues associated with the spatial operator $\hat{A}_{D}$ defined in Eq. \eqref{laplace_Beltrame}.

In order to use Eqs. \eqref{sol_time} and \eqref{eigen1} to calculate the generalized zeta function in $(D+1)$ dimensions, let us consider Eq. \eqref{zeta} in the form \cite{Hawking1977, Elizalde1994book, Kirsten:2010zp, Nesterenko:2004gzu, aleixo2021thermal}
\begin{equation}\begin{split}\label{zeta_trace}
\zeta_{N}(s) = \frac{1}{\Gamma(s)}\sum_{n=-\infty}^{\infty}\int_0^\infty \xi^{s-1}e^{-\left(\frac{2\pi n}{\beta}\right)^2\xi}\text{Tr}\left[e^{-\xi\hat{A}_{D}}\right]d\xi,
\end{split}\end{equation}
where the trace of the operator $\hat{A}_{D}$, as indicated above, is given by
\begin{equation}\begin{split}\label{heat_kernel0}
K_{D}(\xi)=\text{Tr}\left[e^{-\xi\hat{A}_{D}}\right] = \sum_{j}e^{-\Omega_j\xi}.
\end{split}\end{equation}
This expression also defines what is commonly known as heat kernel. Moreover, associated with the above expression there is also a very useful expansion for small $\xi$ which gives information about the ultraviolet divergencies that usually appear in the calculation of the vacuum energy. Such heat kernel expansion may be written as \cite{Elizalde1994book, Kirsten:2010zp, Nesterenko:2004gzu, bordag2009advances}
\begin{equation}\label{heat_expl}
        K_D(\xi)  = \frac{e^{-m^2\xi}}{(4\pi\xi)^{\frac{D}{2}}}\sum_{p=0}^{\infty}C_{\frac{p}{2}}\xi^{\frac{p}{2}} + \text{ES},
\end{equation}
where $C_{\frac{p}{2}}$ are the heat kernel coefficients and `ES' stands for exponentially suppressed terms in the calculation of the integral \eqref{zeta_trace}. The expressions in Eqs. \eqref{heat_kernel0} and \eqref{heat_expl} will play a very important role in our analysis below.

The generalized zeta function in \eqref{zeta_trace} leads to the following expression:
\begin{equation}\begin{split}\label{zeta_temp}
  \zeta_{N}(s)=\frac{\beta}{\sqrt{4\pi}\Gamma(s)}\left\{\Gamma(s-1/2)\zeta_D(s-1/2) + 2\sum_{n=1}^\infty\int_0^\infty \xi^{s-\frac{3}{2}}e^{-\frac{(n\beta)^2}{4\xi}}\text{Tr}\left[e^{-\xi\hat{A}_D}\right]d\xi\right\},
\end{split}\end{equation}
where $\zeta_D(s-1/2) $ is obtained from the term $n=0$ in Eq. \eqref{zeta_trace} . As a consequence, the connection with the partition function $Z$ is written as
\begin{equation}\begin{split}\label{partition}
\ln Z = \frac{1}{2}\zeta'_{N}(0) + \frac{1}{2}\ln\left(\frac{\pi\mu^2}{4}\right)\zeta_{N}(0),
\end{split}\end{equation}
where prime indicates derivation with respect to the argument $s$ of the zeta function \eqref{zeta_temp}. The parameter $\mu$ has dimension of mass and arises as a mesure on the space of field functions in the path integral formulation adopted \cite{Elizalde1994book, aleixo2021thermal}. 

Furthermore, the free energy follows from the partition function above and is given by \cite{Elizalde1994book, Kirsten:2010zp, Nesterenko:2004gzu}
\begin{equation}\begin{split}\label{CasimirEnDenZetaMethod}
    F &= -\frac{1}{\beta}\ln Z  \\
    &= \frac{1}{2}\zeta_D(-1/2) - \frac{\bar{C}_{\frac{N}{2}}}{2(4\pi)^{\frac{N}{2}}}\ln(M^2) - 
  \frac{1}{\sqrt{4\pi}}\sum_{n=1}^\infty\int_0^\infty \xi^{-\frac{3}{2}}e^{-\frac{(n\beta)^2}{4\xi}}\text{Tr}\left[e^{-\xi\hat{A}_D}\right]d\xi,
\end{split}\end{equation}
where $M=\frac{\sqrt{\pi}e^2\mu}{8}$ and the coefficients $\bar{C}_{\frac{N}{2}}$ are given by Eq. \eqref{coef1}, also related to the heat kernel coefficients defined in Eq. \eqref{heat_expl}. While the third term in the r.h.s. of Eq. \eqref{CasimirEnDenZetaMethod} makes possible to obtain temperature corrections to the free energy, the first two terms, by cautiously adopting an adequate renormalization procedure, provide the vacuum energy at zero temperature \cite{Elizalde1994book, Kirsten:2010zp}. In order to better see that, let us write  
\begin{eqnarray}\label{regularized}
  E_0(D) = E_0^{\text{fin}}(D)  - \frac{\bar{C}_{\frac{N}{2}}}{2(4\pi)^{\frac{N}{2}}}\ln(M^2), 
\end{eqnarray}
where 
\begin{equation}\label{ren11}
        E_0^{\text{fin}}(D) = \lim_{s\rightarrow 0}\left[E_0(D,s) - E_{0}^{\text{div}}(D,s)\right]
\end{equation}
is the finite part of the first term in the r.h.s. of Eq. \eqref{CasimirEnDenZetaMethod}. This term can be regularized by defining 
\begin{equation}\label{zetaReg}
        E_0(D,s) =  \frac{1}{2}\zeta_D(s-1/2).
\end{equation}
Furthermore, the divergent contribution indicated in Eq. \eqref{ren11} can be obtained by using the heat kernel expansion \eqref{heat_expl} in the calculation of the regularized expression\footnote{See Refs. \cite{bordag2009advances, Bordag:1998rf} for the case $D=3$.} in Eq. \eqref{zetaReg}. This is achieved by considering the term $n=0$ in Eq. \eqref{zeta_trace}, which gives
\begin{equation}\label{expansion1}
       E_0(D,s) = \frac{1}{2(4\pi)^{\frac{D}{2}}\Gamma\left(s-\frac{1}{2}\right)}\sum_{p=0}^{\infty}\left[\frac{\bar{C}_p}{s-\frac{N}{2}+p} + \frac{\bar{C}_{\frac{2p+1}{2}}}{s-\frac{N}{2}+\frac{2p+1}{2}}\right],
       \end{equation}
with
\begin{equation}\label{coef1}
 \bar{C}_p = \sum_{d=0} ^{p}\frac{(-1)^d}{d!}C_{p-d}m^{2d},\qquad\qquad \bar{C}_{\frac{2p+1}{2}} = \sum_{d=0} ^{p}\frac{(-1)^d}{d!}C_{\frac{2p+1-2d}{2}}m^{2d},
\end{equation}
where we have also used a Taylor expansion, in powers of mass, for $e^{-m^2\xi}$. A carefully analysis of Eq. \eqref{expansion1} shows that the divergent contributions when $s\rightarrow 0$ come from the terms $p=\frac{D+1}{2}$ for odd $D$, and $p=\frac{D}{2}$ for even $D$. Thus,
\begin{equation}\label{pole1}
E^{\text{div}}_0(D,s) = -\frac{1}{2(4\pi)^{\frac{N}{2}}}\frac{\bar{C}_{\frac{N}{2}}}{s},
\end{equation}
where
\begin{equation}\label{coef1}
\bar{C}_{\frac{N}{2}} = \sum_{d=0} ^{p}\frac{(-1)^d}{d!}C_{\frac{N-2d}{2}}m^{2d}.
\end{equation}
%
%
%
%

In the massive case there can be contributions in Eq. \eqref{regularized}, coming from Eq. \eqref{coef1}, that are proportional to the mass with positive powers. As a consequence, these contributions must also be subtracted from Eq. \eqref{regularized} in order to obtain a renormaized expression for the vacuum energy. In this case, the latter should satisfy the normalization condition \cite{bordag2009advances, Bordag:1998rf}
\begin{equation}\label{ren1}
        \lim_{m\rightarrow\infty} E_0^{\text{ren}}(D) = 0,
\end{equation}
which is plausible since in the limit for large masses we should not expect to have a nonzero vacuum energy, which is a quantum effect.

%

As to the operator $\hat{A}_D$ in Eq. \eqref{CasimirEnDenZetaMethod}, it satisfies the eigenvalue equation
\begin{equation}\begin{split}\label{D_EVE}
\hat{A}_D\varphi_{j}=\Omega_{j}\varphi_{j},
\end{split}\end{equation}
where $\Omega_{j}$ is the set of eigenvalues associated with the spatial momenta, as pointed out in Eq. \eqref{eigen1}.

For spacetimes where the operator $\hat{A}_D$ presents explicitly and well defined eigenvalues the generalized zeta function \eqref{zeta_trace} is in geral not difficult to calculate. However, in cases where it is considered curved spacetimes or spacetimes with nontrivial topology, we should use a local approach to calculate the heat kernel \cite{Hawking1977}. This is achieved by considering the local heat kernel definition, that is,
\begin{equation}\label{HK}
    K_D(w,w',\xi) = \sum_{j} e^{-\Omega_{j} \xi}\varphi_{j}(w)\varphi^{*}_{j}(w'),
\end{equation} 
where $w=(x^1,x^2,...,x^D)$ stands for $D$ spatial coordinates. The local heat kernel above satisfies the heat kernel equation
\begin{equation}\label{hk_equation}
    \frac{\partial}{\partial\xi}K_D(w,w',\xi) + A_DK_D(w,w',\xi) = 0
\end{equation} 
with initial condition 
\begin{equation}\label{lhk_ic}
K_D(w,w',0)=\delta^D(w-w').
\end{equation} 

The heat kernel in $D$ spatial dimentions is, thus, obtained in terms of the local heat kernel as follows 
\begin{equation}\label{lhk_ic1}
K_D(\xi) = \int\sqrt{|g^{(D)}|}K_D(w,w,\xi)d^Dw,
\end{equation} 
where we have considered the normalization condition 
\begin{equation}\label{lhk_ic}
\int \sqrt{|g^{(D)}|}\varphi_{j}(w)\varphi^{*}_{j'}(w)d^Dw = \delta_{j,j'}.
\end{equation} 
The delta symbol in the r.h.s. of the above expression is understood as Kronecker delta or Dirac delta function, depending on whether $j$ is discrete or continuous and $|g^{(D)}|$ is the determinant of the spatial part of the metric $g_{\mu\nu}$. Therefore, once obtained the heat kernel by performing the integration in Eq. \eqref{lhk_ic1}, we can in principle calculate the corresponding generalized zeta function.

%

\section{Vacuum energy, temperature corrections and heat kernel coefficients in a quasiperiodically identified conical spacetime}
\label{sec3}
%
A conical spacetime codifies the geometrical structure associated with linear topological defects, such as cosmic strings predicted to exist in the Universe by some extensions of the Standard Model of Particle Physics \cite{hindmarsh, VS} and in String Theory \cite{Copeland:2011dx, Hindmarsh:2011qj}. This defect also exists in condensed matter systems and it is known as disclination \cite{Katanaev:1992kh}.

The line element describing a conical defect in a $(D+1)$-dimensional Euclidean spacetime in cylindrical coordinates is given by 
\begin{equation}\label{stringLineElement}
    ds^2 = -d\tau^2 - dr^2 - r^2 d\phi^2 - dz^2 - \sum_{j=4}^{D}\left(dx^j\right)^2,
\end{equation}
where $\tau = it$ is the imaginary time, $D\ge 4$, $r\geq 0$, $0 \leq \phi \leq 2\pi/q$ and $-\infty \leq (t,z, x^j) \leq\infty$, for $j=4,...,D$. The quantity $\Delta\phi = 2\pi - 2\pi/q$ gives us the planar angle deficit (or excess) caused by the presence of the conical space time. Note that the conical defect is assumed to be on the ($D -2)$-dimensional hypersurface $r=0$. In the usual four dimensions, the parameter $q\ge 1$ is proportional to the linear mass density of the cosmic string $\mu_0$ by $q^{-1} = 1 - 4G\mu_0$, where $G$ is the Newton's gravitational constant \cite{MotaDispiration, klecioQuasiPNanotubes}. In condensed matter, on the other hand, the parameter $q$ characterizes a disclination in solids possessing a spin structure and can assume values such that $q \geq 0$.  

We wish to consider the scalar field's quantum modes propagating in the $(D+1)$-dimensional Euclidean spacetime described by the line element in Eq. \eqref{stringLineElement}. The scalar field is also required to obey a quasiperiodic condition given by
\begin{equation}\label{quasiPCondition}
    \varphi\left(r,\phi,z, x^j\right) = e^{-2\pi i\alpha}\:\varphi\left(r,\phi+2\pi/q,z, x^j\right), 
\end{equation}
where $0 \leq \alpha <1$ is a constant that regulates the phase angle in Eq. \eqref{quasiPCondition}. The special case where $\alpha = 0$ corresponds to the periodic condition, while the special case with $\alpha = 1/2$ corresponds to the antiperiodic condition.

From the eigenvalue equation \eqref{D_EVE}, and taking into consideration the line element \eqref{stringLineElement}, the spatial part of the scalar field \eqref{sol_time} can be obtained by solving the equation
\begin{equation} \label{EigenvalueEqCilindrical}
    \left[-\frac{1}{r}\frac{\partial}{\partial r}\left(r\frac{\partial}{\partial r}\right) -\frac{1}{r^2}\frac{\partial^2}{\partial \phi^2}-\frac{\partial^2}{\partial z^2}+ \sum_{j=4}^{D}\frac{\partial^2}{\partial x^{j2}}+m^2 \right]\varphi_j(w) = \lambda_j\varphi_j(w),
\end{equation}
where $j$ stands for the spatial quantum modes.

The complete set of normalized solutions for the eigenvalue equation \eqref{EigenvalueEqCilindrical} under the quasiperiodic condition \eqref{quasiPCondition} on a conical spacetime is given by \cite{MotaDispiration, klecioQuasiPNanotubes, deFarias:2021qdg}
\begin{equation}\label{nonNormalizedGenSolution}
    \varphi_j\left(\mathbf{r}\right) = \left[\frac{q\eta}{(2\pi)^{D-1}}\right]^{\frac{1}{2}} e^{i{\bf p}\cdot{\bf r}_{\parallel} + i\nu z+iq(\ell+\alpha)\phi} J_{q|\ell+\alpha|}(\eta r),
\end{equation} 
where ${\bf r}_{\parallel}$ and ${\bf p}$ stand, respectively, for the coordinates of the extra dimensions and their corresponding momenta,  $J_\mu(x)$ is the Bessel function of the first kind and $j = \left(p,\nu,\eta,\ell\right)$ is the set of quantum numbers. In this case, the eigenvalues are found to be
\begin{equation}\label{eigenvalues}
    \Omega_j = p^2+\nu^2 + \eta^2 + m^2. 
\end{equation}

Now we can use Eqs. \eqref{nonNormalizedGenSolution} and \eqref{eigenvalues} with Eq. \eqref{HK}. The sum for the set of quantum numbers of the problem takes the form
\begin{equation}
    \sum_j = \int d^{D-3}p\int_{-\infty}^{\infty} d\nu \int_0^\infty d\eta \sum_{\ell=-\infty}^{\infty}.
    \label{SDef}
\end{equation}
Consequently, the local heat kernel defined in Eq. \eqref{HK} is written as
\begin{equation} \begin{split} \label{Heat1}
    K _D(w,w',\xi) = \frac{q}{(2\pi)^{D-1}}\sum_{j}e^{-\Omega_{j}\xi}e^{i\ell\Delta\phi + i\nu\Delta z + i{\bf p}\cdot\Delta{\bf r}_{\parallel}}\eta J_{q|\ell+\alpha|}(\eta r)J_{q|\ell+\alpha|}(\eta' r),
\end{split} \end{equation}
where $\Delta\phi=\phi - \phi'$, $\Delta z=z-z'$ and $\Delta{\bf r}_{\parallel}= {\bf r}_{\parallel} - {\bf r}_{\parallel}' $. The type of calculation involved in the above expression has been developed in several works, as for instance in Refs. \cite{MotaDispiration, klecioQuasiPNanotubes, deFarias:2021qdg, BezerradeMello:2014phm}. In this sense, by using the method adopted in these works we found
\begin{equation} \begin{split} \label{HeatKernel2pts}
    K_D (w,w',\xi) = & \frac{e^{-m^2\xi}}{(4\pi\xi)^{\frac{D}{2}}}\Biggl\{\sum_{\ell} \frac{e^{i\alpha(2\pi \ell - q\Delta\phi)}}{e^{\frac{\Delta R_\ell^2}{4\xi}}} - \frac{q}{2\pi i}\sum_{b=\pm 1} be^{ibq\alpha\pi} \int_0^\infty dy \: \frac{\cosh[qy(1-\alpha)] - \cosh(qy\alpha)e^{-iq(\Delta\phi +b\pi)}}{e^{ \frac{\Delta R_y ^2}{4\xi}}\left[\cosh (qy)-\cos(q\Delta\phi+bq\pi)\right]} \Biggl\},
\end{split} \end{equation}
where the index $\ell$ must obey the restriction \cite{BezerradeMello:2014phm}
\begin{equation}
-\frac{q}{2} + \frac{q\Delta\phi}{2\pi}\le \ell \le \frac{q}{2} + \frac{q\Delta\phi}{2\pi}.
\label{restriction}
\end{equation}
Moreover, we have also introduced the definition
    \begin{eqnarray}
        \Delta R_\ell^2 &=& \Delta z^2 + \Delta r_{\parallel}^2+r^2+r'^2-2rr'\cos\left(\frac{2\pi\ell}{q} - \Delta\phi \right)\nonumber\\
          \Delta R_y^2 &=& \Delta z^2 + \Delta r_{\parallel}^2+r^2+r'^2+2rr'\cosh(y).
    \end{eqnarray}

Upon taking the coincidence limit $w'\rightarrow w$ in Eq. \eqref{HeatKernel2pts} we have
\begin{equation}\label{heatKernelCoincid}
        K_D(w,w,\xi) = K_D^{\text{E}}(w,w, \xi)  + \frac{e^{-m^2\xi}}{(4\pi\xi)^{\frac{D}{2}}}\Biggl\{2\sum_{\ell=1}^{[q/2]}\cos(2\pi\ell\alpha)e^{-\frac{(2rs_\ell)^2}{4\xi}} -\frac{q}{\pi}\int_0^\infty dy \:M(y,\alpha,q)\: e^{-\frac{(2rs_y)^2}{4\xi}}\Biggl\},
\end{equation}
where $s_{\ell}=\sin(\pi\ell/2)$, $s_{y}=\cosh(y/2)$ and $[q/2]$ stands for the integer part of $q/2$ and in the case it is an integer the corresponding term in the sum should be taken with the coefficient 1/2. Note also that
\begin{equation} \label{Myalphaq}
    M(y,\alpha,q) = \frac{\cosh[qy(1-\alpha)]\sin(q\pi\alpha)+\cosh(qy\alpha)\sin[q\pi(1-\alpha)]}{\cosh(qy)-\cos(q\pi)}.
\end{equation}
The first term in the r.h.s. of Eq. \eqref{heatKernelCoincid} is the Euclidean local heat kernel contribution that comes from the term $\ell=0$ in the sum present in Eq. \eqref{HeatKernel2pts}. This contribution is given by
\begin{equation}\label{EuclideanHK}
    K_D^{\text{E}}(w,w, \xi) = \frac{1}{(4\pi\xi)^\frac{D}{2}}e^{-m^2\xi}.
\end{equation}

By using \eqref{lhk_ic1} and \eqref{heatKernelCoincid} the heat kernel is found to be
\begin{equation}\label{heatglobalstringQ}
        K_D(\xi) = \frac{e^{-m^2\xi}}{(4\pi\xi)^{\frac{D}{2}}}C_0 + \frac{\xi e^{-m^2\xi}}{(4\pi\xi)^{\frac{D}{2}}}C_1,
\end{equation}
%
where $C_0$ and $C_1$ are the heat kernel coefficients obtained by comparison of the above expression with the heat kernel expansion in Eq. \eqref{heat_expl}. These nonzero coefficients are given by
\begin{equation}\label{C0}
   C_0 = V_{D}
 \end{equation}
and
\begin{eqnarray}\label{pole_ZF}
   C_1 &= &h(q,\alpha)V_{D-2}\nonumber\\
 &=&\Biggl\{ \sum_{\ell=1}^{[q/2]}\frac{\cos(2\pi\ell\alpha)}{s_{\ell}^2} -\frac{q}{2\pi}\int_0^\infty dy \:\frac{M(y,\alpha,q)}{s_y^2}\Biggl\}\frac{2\pi}{q}V_{D-2},
 \end{eqnarray}
where $V_D$ and $V_{D-2}$ are infinite volumes. The $C_0$ coefficient is associated with the Euclidean heat kernel in the first term in the r.h.s. of Eq. \eqref{heatglobal}. The latter provides a divergent contribution to the zeta function $\zeta_{D}(s-1/2)$ that can be obtained from the term $n=0$ in \eqref{zeta_trace}. The coefficient $C_1$ shows a dependence on the parameter $q$ and also on $\alpha$. Thus, it exists as a consequence of the topology of the conical spacetime as well as of the quasiperiodic condition. It is associated with additional ultraviolet divergencies that should be subtracted as we shall see below. We have checked that for $\alpha=0$ the coefficient $C_1$ above recovers known results found in Refs. \cite{cognola:1993qg, Fursaev:1993qk}, as it is shown in Eq. \eqref{pole_ZF_q}. In contrast, for $q=1$, we have only a contribution due to the quasiperiodic condition that comes from the second term in the r.h.s. of Eq. \eqref{pole_ZF}. Thus, here, we have generalized the results of Refs. \cite{cognola:1993qg, Fursaev:1993qk} by including the quasiperiodicity. 

\begin{figure}[h]
    \includegraphics[scale=0.35]{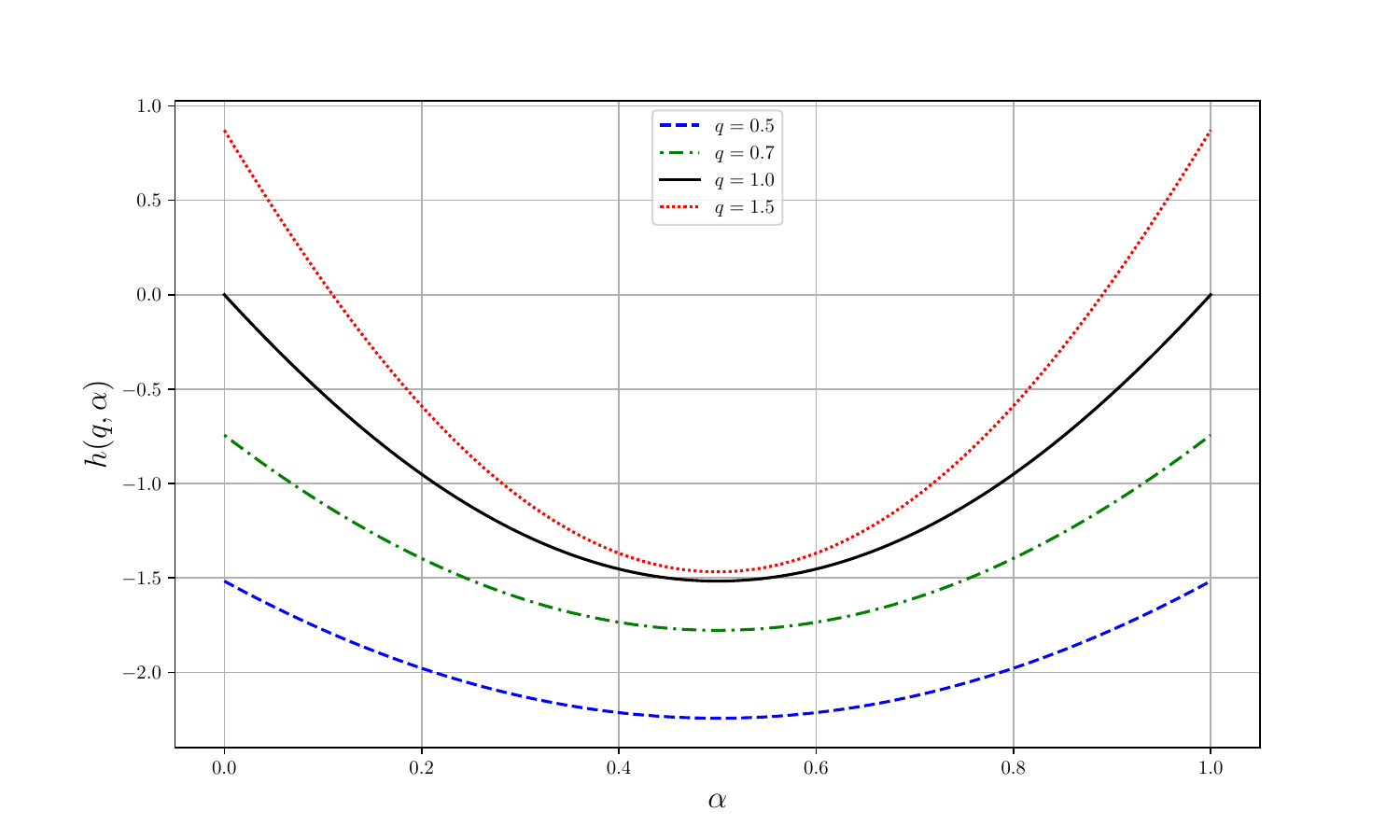}
     \includegraphics[scale=0.35]{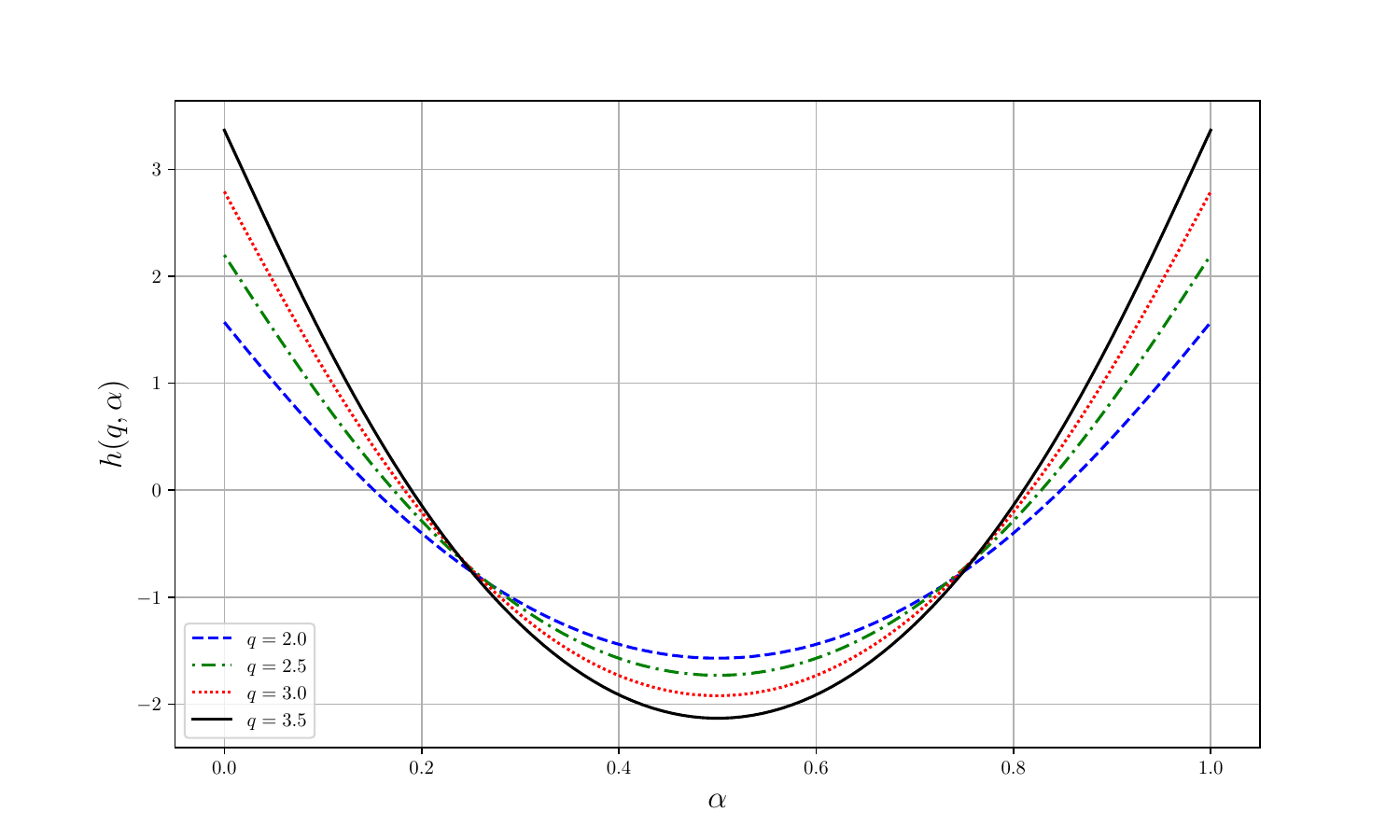}
    \caption{Function $h(q,\alpha)$ plotted in terms of $\alpha$, for several values of $q$.}
    \label{HKC1}
\end{figure}

In Fig.\ref{HKC1} we have plotted, in terms of $\alpha$, the function $h(q,\alpha)$ in Eq. \eqref{pole_ZF} which carries the nontrivial topology information of the heat kernel coefficient  $C_1$. From the plots, we can see that the minimum value for $C_1$ is always when $\alpha=1/2$, the antiperiodic case. The most interesting information in the plots, though, is that for $q> 1$ there are some values of $\alpha$ such that the coefficient $C_1$ vanishes. This eliminates the divergent contribution in Eq. \eqref{energytotal1} due to the nontrivial topology of the spacetime, leaving only the usual Euclidean (Minkowski) divergence related to the coefficient $C_0$.

From \eqref{heatglobal} and \eqref{zeta_trace} we have the generalized zeta function as given by
\begin{eqnarray}\label{genZeta}
        \Gamma(s-1/2)\zeta_D(s-1/2) &=&  \frac{m^{-2\left(s-\frac{D+1}{2}\right)}}{(4\pi)^{\frac{D}{2}}}\Gamma\left(s-\frac{D+1}{2}\right)C_0\nonumber\\
        &+& \frac{m^{-2\left(s-\frac{D-1}{2}\right)}}{(4\pi)^{\frac{D}{2}}}\Gamma\left(s-\frac{D-1}{2}\right)C_1,
\end{eqnarray}
from which the regularized vacuum energy can be obtained by using Eq. \eqref{zetaReg}. Note that the expression above depends on the spatial dimension, that is, if $D$ is even we have a straightforward finite result, after taking the limit $s\rightarrow 0$, while if $D$ is odd there will be divergent contributions, along with finite ones. In this case, the divergent contributions to the vacuum energy is written as
\begin{equation}\label{energytotal1}
     2(4\pi)^{\frac{D+1}{2}}E_{0}^{\text{div}}(D,s) = - \frac{1}{s}\left[\frac{(-1)^{\frac{D+1}{2}}m^{D+1}}{\Gamma\left(\frac{D+3}{2}\right)}C_0 + \frac{(-1)^{\frac{D-1}{2}}m^{D-1}}{\Gamma\left(\frac{D+1}{2}\right)}C_1\right].
\end{equation}
By comparing the expression above with Eq. \eqref{pole1}, it is clear that 
\begin{equation}\label{Dpuls1}
    \bar{C}_{\frac{N}{2}}=\left[\frac{(-1)^{\frac{D+1}{2}}m^{D+1}}{\Gamma\left(\frac{D+3}{2}\right)}C_0 + \frac{(-1)^{\frac{D-1}{2}}m^{D-1}}{\Gamma\left(\frac{D+1}{2}\right)}C_1\right]
\end{equation}
and, as a consequence, the second term in the r.h.s. of Eq. \eqref{regularized} is completely determined. Furthermore, still for odd $D$, upon taking the limit $s\rightarrow 0$ of Eq. \eqref{genZeta} there will be finite contributions that logarithmically depend on the mass $m$, as well as on the mass with positive powers. Thus, after subtracting the divergent contribution \eqref{energytotal1} according to the prescription \eqref{ren11}, from Eq. \eqref{regularized}, we obtain for odd $D$ that
\begin{eqnarray}\label{energytotal}
 2(4\pi)^{\frac{D+1}{2}}E_{0}(D) &=&\frac{(-1)^{\frac{D+3}{2}}m^{D+1}}{\Gamma\left(\frac{D+3}{2}\right)}C_0\left[\frac{1}{\Gamma\left(\frac{D+3}{2}\right)} - \ln\left(\frac{4M^2}{m^2}\right)\right]\nonumber\\
 &+& \frac{(-1)^{\frac{D+1}{2}}m^{D-1}}{\Gamma\left(\frac{D+1}{2}\right)}C_1\left[\frac{1}{\Gamma\left(\frac{D+1}{2}\right)} - \ln\left(\frac{4M^2}{m^2}\right)\right],
\end{eqnarray}
where the logarithmic mass terms from the finite contribution and the second term in the r.h.s. of Eq. \eqref{regularized} have been put together. Note that the massless limit of the above expression vanish. 

Although now the vacuum energy in Eq. \eqref{energytotal} is finite, it still depends on the mass scale $M$ due to the ambiguity in the r.h.s. of Eq. \eqref{regularized}. Moreover, the expression above increases without bounds with the mass $m$, in contrast with the fact that the vacuum energy is purely a quantum phenomenon and, hence, should vanish in the limit of very large masses. In this sense, the additional requirement encoded in Eq. \eqref{ren1} should be satisfied in order to obtain a renormalized vacuum energy \cite{bordag2009advances, Bordag:1998rf}. This means that, we must also subtract all the terms in the vacuum energy that increase with the mass $m$. In fact, a brief inspection of Eq. \eqref{energytotal} for odd $D$ and of Eq. \eqref{genZeta} for even $D$ shows that all terms present in theses expressions increase with the mass and are to be subtracted. Of course, this leads to a zero renormalized vacuum energy in the quasiperiodically identified conical spacetime considered, that is,
\begin{equation}\label{RenVE}
  E^{\text{ren}}_0(D) = 0, 
 \end{equation}
which trivially satisfies the requirement in Eq. \eqref{ren1}. The result above for the renormalized vacuum energy has not been pointed out in the literature so far for the system analyzed in this section, including the periodic case $\alpha=0$. Moreover, a similar result was obtained in Refs. \cite{Khusnutdinov:1998tf}, where the authors considered the scalar vacuum energy in the cosmic string spacetime under Dirichlet boundary condition.

For $\alpha =0$ and integer values of $q$, we can make the following replacement in the coefficient $C_1$ in Eq. \eqref{pole_ZF}:
\begin{equation}\label{replacement}
  \sum_{\ell=1}^{[q/2]}\rightarrow\frac{1}{2}\sum_{\ell=1}^{q-1}
   \end{equation}
with
\begin{equation} \label{Myq}
    M(y,0,q) = \frac{\sin(q\pi)}{\cosh(qy)-\cos(q\pi)}.
\end{equation}
If $q$ is an integer, it is obvious that, by using \eqref{replacement}, the heat kernel coefficient $C_1$ for $\alpha=0$ reduces to
\begin{eqnarray}\label{pole_ZF_q}
   C_1 &=&\frac{2\pi}{q}V_{D-2}\frac{1}{2}\sum_{\ell=1}^{q-1}\frac{1}{s_{\ell}^2}\nonumber\\
   &=& V_{D-2}\frac{\pi}{3}\left(q-\frac{1}{q}\right),
 \end{eqnarray}
 which is in essence the result found in Refs. \cite{cognola:1993qg, Fursaev:1993qk}. The expression above is an analytic function of $q$ and, by analytic continuation, should be valid for any $q$.

On the other hand, in the case $\alpha=1/2$, we can still make use of Eq. \eqref{replacement} but only if $q$ is an even integer. Hence, by noting that $M(y,1/2,q) =0$ from Eq. \eqref{Myalphaq}, we have 
\begin{eqnarray}\label{pole_ZF_beta_half}
   C_1 &=&\frac{2\pi}{q}V_{D-2}\frac{1}{2}\sum_{\ell=1}^{q-1}\frac{(-1)^{\ell}}{s_{\ell}^2}\nonumber\\
   &=&-V_{D-2}\frac{\pi}{6}\left(q+\frac{2}{q}\right).
 \end{eqnarray}
 This result although has been obtained by considering even integer values of $q$, it is by analytic continuation in fact valid for any value of $q$. In particular, for $q=1$, we have $C_1=-\frac{\pi}{2}V_{D-2}$. Note that both expressions in Eqs. \eqref{pole_ZF_q} and \eqref{pole_ZF_beta_half} are in agreement with the plots shown in Fig.\ref{HKC1}.

%
\subsection{Temperature corrections}
Let us now turn to the calculation of possible temperature corrections to the renormalized vacuum energy. This is achieved by making use of the third term in the r.h.s. of Eq. \eqref{CasimirEnDenZetaMethod}, i.e.,
\begin{equation}\begin{split}\label{TemExp}
    F_T = -  \frac{1}{\sqrt{4\pi}}\sum_{n=1}^\infty\int_0^\infty \xi^{-\frac{3}{2}}e^{-\frac{(n\beta)^2}{4\xi}}\text{Tr}\left[e^{-\xi\hat{A}_D}\right]d\xi.
\end{split}\end{equation}
Thus, by taking into consideration Eq. \eqref{heatglobalstringQ} the above expression provides
\begin{equation}\begin{split}\label{Temp_corr1}
    F_{T}= - \frac{2^{\frac{D+3}{2}}m^{D+1}C_0}{(4\pi)^{\frac{D+1}{2}}}\sum_{n=1}^{\infty}f_{\frac{D+1}{2}}(mn\beta) - \frac{2^{\frac{D+1}{2}}m^{D-1}C_1}{(4\pi)^{\frac{D+1}{2}}}\sum_{n=1}^{\infty}f_{\frac{D-1}{2}}(mn\beta),
\end{split}\end{equation}
for the massive scalar field case. The function $f_{\mu}(z)$ is defined in terms of the Macdonald function $K_{\mu}(z)$ as
\begin{equation}\begin{split}\label{funF}
 f_{\mu}(z) = \frac{K_{\mu}(z)}{z^{\mu}}.
\end{split}\end{equation}

The massless temperature correction is obtained from Eq. \eqref{Temp_corr1} by noting the following limit \cite{abramowitz, gradshteyn2000table}
\begin{equation}\begin{split}\label{limite}
\lim_{z\rightarrow 0}z^{\mu}K_{\mu}(nz)=\frac{\Gamma(\mu)}{2}\left(\frac{2}{n}\right)^{\mu}.
\end{split}\end{equation}
%
Hence, this gives
\begin{equation}\begin{split}\label{Temp_corr2}
    F_{T}= - \frac{2^{D+1}C_0}{(4\pi)^{\frac{D+1}{2}}\beta^{D+1}}\Gamma\left(\frac{D+1}{2}\right)\zeta_{\text{R}}(D+1) - \frac{2^{D-1}C_1}{(4\pi)^{\frac{D+1}{2}}\beta^{D-1}}\Gamma\left(\frac{D-1}{2}\right)\zeta_{\text{R}}(D-1),
        \end{split}\end{equation}
where $\zeta_{\text{R}}(s)$ is the Riemann zeta function.

In particular, for $D=3$, we find
\begin{equation}\begin{split}\label{Temp_corr3}
    F_{T}= -\frac{\pi^2}{90}C_0\frac{(k_BT)^4}{(\hbar c)^4} - \frac{1}{24}C_1\frac{(k_BT)^2}{(\hbar c)},
        \end{split}\end{equation}
for the massless scalar field case. Note that the first term in the r.h.s. is the scalar black body radiation contribution and it is related to the heat kernel coefficient $C_0$, in general associated with the Minkowski spacetime \cite{Bezerra:2011nc, PhysRevD.83.104042}. On the other hand, the second term is a contribution due to the spacetime conical topology and quasiperiodicity. This term, as we can see, it is related to the heat kernel coefficient $C_1$. As we have learned, nonzero heat kernel coefficients indicate the existence of divergences and, as such, both terms in Eqs. \eqref{Temp_corr1}, \eqref{Temp_corr2} and \eqref{Temp_corr3} should be subtracted. With this, we obtain a zero renormalized temperature correction contribution, i.e.,
 \begin{equation}\label{RenFE}
  F_T^{\text{ren}} = 0.
 \end{equation}

In the massless case, and for $D=3$, it has also been argued that terms of the type 
 \begin{eqnarray}
\alpha_0\frac{(k_BT)^4}{(\hbar c)^3},\qquad\qquad\alpha_1\frac{(k_BT)^3}{(\hbar c)^2}, \qquad\qquad\alpha_2\frac{(k_BT)^2}{\hbar c},
\label{QNterms}
\end{eqnarray}
must be subtracted in the renormalization process in order to obtain a correct classical contribution in the high-temperature limit \cite{Bezerra:2011nc, PhysRevD.83.104042, Mota:2022qpf}. In other words, the expressions above are all of quantum nature and should not dominate in the high-temperature limit. In Refs. \cite{Bezerra:2011nc, PhysRevD.83.104042, Mota:2022qpf} it has also been pointed out that the parameters $\alpha_0$, $\alpha_1$ and $\alpha_2$ are related to the heat kernel coefficients and we have indeed shown here that this is in fact the case, at least for $\alpha_0$ and $\alpha_2$. Upon comparing Eqs. \eqref{QNterms} and \eqref{Temp_corr3} we conclude that 
 \begin{eqnarray}
\alpha_0=-\frac{\pi^2}{90}C_0 ,\qquad\qquad\alpha_2=- \frac{1}{24}C_1.
\label{alphas}
\end{eqnarray}
Of course, the discussion above is to say, in other words, that terms containing any heat kernel coefficient must be subtracted. 

The main important expressions derived in this section have been the heat kernel two-point function, the corresponding heat kernel coefficients, the vacuum energy and its temperature corrections. The heat kernel two-point function \eqref{heatKernelCoincid} in the coincidence limit retains information about the existing divergences associated with the system configuration. In our case, there is a divergence brought upon by the Minkowski spacetime flat nature, normally present, and also a divergence brought upon by the topological conical structure of the quasiperiodically spacetime. The mathematical objects responsible for identifying these divergences are the heat kernel coefficients, in our case given by Eq. \eqref{C0} and \eqref{pole_ZF} and which are present in Eq. \eqref{heatglobalstringQ}. Thus, in order to obtain meaningful physical interpretations for the vacuum energy and its temperature corrections free of divergences one needs to subtract the terms containing the mentioned heat kernel coefficients in a well defined renormalized process, as it has been done here. The result, as we have seen, it is zero for both the renormalized vacuum energy and temperature corrections.

In the next section we shall analyze, by using the same technique, the vacuum energy and temperature corrections arising as a consequence of the nontrivial topology of a cosmic dispiration, a combination of a conical defect, like a cosmic string or disclination, with another defect known as screw dislocation.

\section{VACUUM ENERGY AND TEMPERATURE CORRECTIONS IN A COSMIC DISPIRATION SPACETIME}
\label{sec4}
%
We want now to consider a spacetime geometry that is a combination of two topological defects, that is, a conical and a screw dislocation, forming what is known as cosmic dispiration spacetime \cite{Galtsov:1993ne, DeLorenci:2003wv, DeLorenci:2002jv}. The line element that describes an idealized ($D+1$)-dimensional spacetime that characterizes this combination, in cylindrical coordinates, is given by \cite{MotaDispiration}
\begin{equation}
ds^2=dt^2-dr^2-r^2d\phi^2-(dz+\kappa d\phi)^2-\sum_{i=4}^{D}(dx^i)^2,
\label{Me}
\end{equation}
where $\kappa$ is a constant parameter associated with the screw dislocation and that has dimension of length, $D>3$ and $(r,\phi,z,x^4,...,x^D)$ are the generalized cylindrical coordinates taking values in the ranges $r\geq 0$, $0\leq\phi\leq \phi_0=2\pi/q$ and $-\infty<(t,z,x^i)<+\infty$, for $i=4,...,D$. The conical defect carries all the properties described below Eq. \eqref{stringLineElement}.

From the eigenvalue equation \eqref{D_EVE}, and taking into consideration the line element \eqref{Me}, the spatial part of the scalar field \eqref{sol_time} can be obtained by solving the equation
\begin{eqnarray}
\left[-\frac{1}{r}\frac{\partial}{\partial r}\left(r\frac{\partial}{\partial r}\right)-\frac{1}{r^2}\left(\frac{\partial}{\partial\phi}-\kappa\frac{\partial}{\partial z}\right)^2-\frac{\partial^2}{\partial z^2}-\sum_{i=4}^{D}\frac{\partial^2}{\partial x^{i2}}+m^2\right]\varphi_j(w) = \Omega_j\varphi_j(w),
\label{KG2}
\end{eqnarray}
where $j$ stands for the spatial quantum modes.

The complete set of normalized solutions for the eigenvalue equation \eqref{KG2} is given by \cite{MotaDispiration}
 \begin{eqnarray}
 \varphi_j\left(\mathbf{r}\right)=\left[\frac{q\eta}{(2\pi)^{D-1}} \right]^{\frac{1}{2}}J_{|q\ell-\kappa\nu|}(\eta r)e^{i\ell q\phi+i\nu z+i{\bf p}\cdot {\bf r_{\parallel}} },
\label{SS2}
\end{eqnarray}
where ${\bf r}_{\parallel}$ and ${\bf p}$ stand, respectively, for the coordinates of the extra dimensions and their corresponding momenta,  $J_\mu(x)$ is the Bessel function of the first kind and $j = \left(p,\nu,\eta,\ell\right)$ is the set of quantum numbers. Note that the eigenvalues are still given by Eq. \eqref{eigenvalues}.

The local heat kernel can be obtained by using Eq. \eqref{HK} and the complete set of normalized solutions \eqref{SS2}, i.e.,
\begin{equation} \begin{split} \label{Heat1}
    K _D(w,w',\xi) = \frac{q}{(2\pi)^{D-1}}\sum_{j}e^{-\Omega_{j}\xi}e^{i(q\ell-\kappa\nu)\Delta\phi + i\nu\Delta Z + i{\bf p}\cdot\Delta{\bf r}_{\parallel}}\eta J_{|q\ell-\kappa\nu|}(\eta r)J_{|q\ell-\kappa\nu|}(\eta r'),
\end{split} \end{equation}
where the sum in $j$ above is defined in Eq. \eqref{SDef} and we have used the coordinate transformation $Z=z+\kappa\phi$ in Eq. \eqref{SS2} (see Ref. \cite{MotaDispiration} for more details). The type of calculation involved in the above expression has been developed in Ref. \cite{MotaDispiration} and, in this sense, by following the same steps we found
\begin{equation} \begin{split} \label{Hkernel2}
    K_D (w,w',\xi) = & \frac{e^{-m^2\xi}}{(4\pi\xi)^{\frac{D}{2}}}\Biggl\{\sum_{\ell} e^{-\frac{\Delta\rho_\ell^2+\Delta Z_n^2}{4\xi}} - \frac{q}{\pi^2}\sum_{k=-\infty}^{\infty}\int_0^\infty dy \; e^{-\frac{\Delta\rho_y^2+\Delta Z_k^2}{4\xi}}M_{k,q}(\Delta\phi, y)  \Biggl\},
\end{split} \end{equation}
where the form of the function $M_{k,q}(\Delta\phi, y)$ has been obtained in Ref. \cite{MotaDispiration} and, in the coincidence limit, that is, for $\Delta\phi=0$, it is given by Eq. \eqref{B5}. The sum in $\ell$ must also obey the restriction in Eq. \eqref{restriction} and 

    \begin{eqnarray}
        \Delta\rho_n^2 &=& \Delta r_{\parallel}^2+r^2+r'^2-2rr'\cos\left(\frac{2\pi\ell}{q} - \Delta\phi \right),\nonumber\\
         \Delta\rho_y^2 &=&  \Delta r_{\parallel}^2+r^2+r'^2+2rr'\cosh(y).
    \end{eqnarray}
Note that we have also defined $\bar{\kappa}=\frac{2\pi\kappa}{q}$ and $\Delta Z_{\ell} = \Delta Z - \bar{\kappa}b$, where $b=\ell$ and $b=k$ in the first and second terms in the r.h.s. of Eq. \eqref{Hkernel2}, respectively. 

The quantity of interest here is the one in Eq. \eqref{Hkernel2} taken in the coincidence limit $w'\rightarrow w$. This gives 

\begin{eqnarray}\label{Hkernel2C}
        K_D(w,w,\xi) &=&K_D^{\text{E}}(w,w, \xi) \nonumber\\
        &+& \frac{e^{-m^2\xi}}{(4\pi\xi)^{\frac{D}{2}}}\Biggl\{2\sum_{\ell=1}^{[q/2]}e^{-\frac{\bar{\kappa}^2\ell^2}{4\xi}}e^{-\frac{r^2s_\ell^2}{\xi}} - \frac{q}{\pi^2}\sum_{k=-\infty}^{\infty}\int_0^\infty dy \; e^{-\frac{\bar{\kappa}^2k^2}{4\xi}}e^{-\frac{r^2s_y^2}{\xi}}M_{k,q}(0, y)\Biggl\},
\end{eqnarray}
with the Euclidean local heat kernel given by Eq. \eqref{EuclideanHK},  $s_{\ell} = \sin(\pi/2)$, $s_y = \cosh(y/2)$ and $[q/2]$ stands for the integer part of $q/2$ and in the case it is an integer the corresponding term in the sum should be taken with the coefficient 1/2 \cite{MotaDispiration}.

Now, the heat kernel is obtained by integrating Eq. \eqref{Hkernel2C} in whole space, providing the following expression:
\begin{eqnarray}\label{heatglobal}
        K_D(\xi) &=& \frac{e^{-m^2\xi}}{(4\pi\xi)^{\frac{D}{2}}}V_D\nonumber\\
        &+& \frac{\xi e^{-m^2\xi}}{(4\pi\xi)^{\frac{D}{2}}}\Biggl\{\sum_{\ell=1}^{[q/2]}\frac{e^{-\frac{\bar{\kappa}^2\ell^2}{4\xi}}}{s_\ell^2} - \frac{q}{2\pi^2}\sum_{k=-\infty}^{\infty}\int_0^\infty dy \; \frac{e^{-\frac{\bar{\kappa}^2k^2}{4\xi}}}{s_y^2}M_{k,q}(0, y)\Biggl\}\frac{2\pi}{q}V_{D-2},
\end{eqnarray}
where
\begin{eqnarray}
M_{k,q}(0, y) = \frac{\left(\frac{q}{2} + k\right)}{\left(\frac{q}{2} + k\right)^2 + \left(\frac{y}{\phi_0}\right)^2}.
\label{B5}
\end{eqnarray}
By comparing Eq. \eqref{heatglobal} with the heat kernel expansion in Eq. \eqref{heat_expl} we can infer that the only nonzero heat kernel coefficients are $C_0=V_D$ and $C_1$ given by Eq. \eqref{HKCD1}. Then, we have
\begin{eqnarray}\label{heatglobal2}
        K_D(\xi) &=& \frac{e^{-m^2\xi}}{(4\pi\xi)^{\frac{D}{2}}}C_0 + \frac{\xi e^{-m^2\xi}}{(4\pi\xi)^{\frac{D}{2}}}C_1\nonumber\\
        &+&\frac{2\pi}{q}V_{D-2}\frac{\xi e^{-m^2\xi}}{(4\pi\xi)^{\frac{D}{2}}}\Biggl\{\sum_{n=1}^{[q/2]}\frac{e^{-\frac{\bar{\kappa}^2n^2}{4\xi}}}{s_n^2} - \frac{q}{2\pi^2}\sideset{}{'}\sum_{k=-\infty}^{\infty}\int_0^\infty dy \; \frac{e^{-\frac{\bar{\kappa}^2k^2}{4\xi}}}{s_y^2}M_{k,q}(0, y)\Biggl\},
\end{eqnarray}
where prime in the sum in $k$ indicates that $k=0$ is excluded. The latter provides the term with the coefficient $C_1$ in the above expression, with

    \begin{eqnarray}\label{HKCD1}
       C_1 &=& -\frac{2\pi V_{D-2}}{q}\int_0^\infty \frac{dy}{\cosh^2(y/2)}\frac{1}{(\pi^2 + y^2)}\nonumber\\
              &\simeq&-\frac{\pi }{3q}V_{D-2}.
    \end{eqnarray}
Note that, numerically, the integral above is approximately given by $1/6$. One interesting aspect about Eq. \eqref{heatglobal} is that in the case $\kappa=0$, the sum in $k$ of the function $M_{k,q}(0, y) $ converges to \cite{MotaDispiration}
\begin{eqnarray}
M_{q}(0, y) &=&\sum_{k=-\infty}^{\infty}M_{k,q}(0,y)\nonumber\\
&=& \frac{ \pi\sin(q\pi)}{\cosh(q y)-\cos(q\pi)}.
\label{B6}
\end{eqnarray}
This allows us to recover as a limiting case the heat kernel \eqref{heatKernelCoincid} for $\alpha=0$, associated with a conical spacetime. Consequently, the heat kernel coefficient $C_1$ becomes \eqref{pole_ZF_q}. However, as we can see from Eq. \eqref{HKCD1}, if we assume $\kappa\neq 0$ the structure of $C_1$ changes as a consequence of the different topology of the spacetime we are considering in this section. Surprisingly, the coefficient above does not depend on $\kappa$, although its form is due to $\kappa\neq 0$.

The next step is to compute the vacuum energy by making use of the zeta function \eqref{zeta_trace} for $n=0$ and Eq. \eqref{zetaReg}. In this case, the zeta function is found to be
\begin{eqnarray}\label{renEdis11}
 \Gamma(s-1/2)\zeta_D(s-1/2) &=& \frac{m^{-2\left(s-\frac{D+1}{2}\right)}}{(4\pi)^{\frac{D}{2}}}\Gamma\left(s-\frac{D+1}{2}\right)C_0+ \frac{m^{-2\left(s-\frac{D-1}{2}\right)}}{(4\pi)^{\frac{D}{2}}}\Gamma\left(s-\frac{D-1}{2}\right)C_1\nonumber\\
&+& \frac{2\pi}{q}V_{D-2}\frac{2^{\frac{D+1-2s}{2}}m^{D-1-2s}}{(4\pi)^{\frac{D}{2}}}\nonumber\\
&\times&\Biggl\{\sum_{\ell=1}^{[q/2]}\frac{f_{\frac{D-1-2s}{2}}(m\bar{\kappa}\ell)}{s_{\ell}^2} - \frac{q}{2\pi^2}\int_0^\infty  \; \frac{dy}{s_y^2}\sideset{}{'}\sum_{k=-\infty}^{\infty}f_{\frac{D-1-2s}{2}}(|m\bar{\kappa}k|)M_{k,q}(0, y)\Biggl\},
 \end{eqnarray}
where the function $f_{\mu}(z)$ has been defined in Eq. \eqref{funF} and the first two terms in the r.h.s. resemble the form of the zeta function in Eq. \eqref{genZeta}, but in this case with $C_1$ given by \eqref{HKCD1}. These terms, as we have argued before, have to be subtracted in order for the renomalized vacuum energy to satisfy the requirement in Eq. \eqref{ren1}. This also includes, in the massive case, the subtraction of the term that contains the coefficient $\bar{C}_{\frac{N}{2}}$ in Eq. \eqref{regularized}. In our case, this coefficient has also the same structure of Eq. \eqref{Dpuls1}, but again with $C_1$ given by \eqref{HKCD1}.

\begin{figure}[h]
    \includegraphics[scale=0.35]{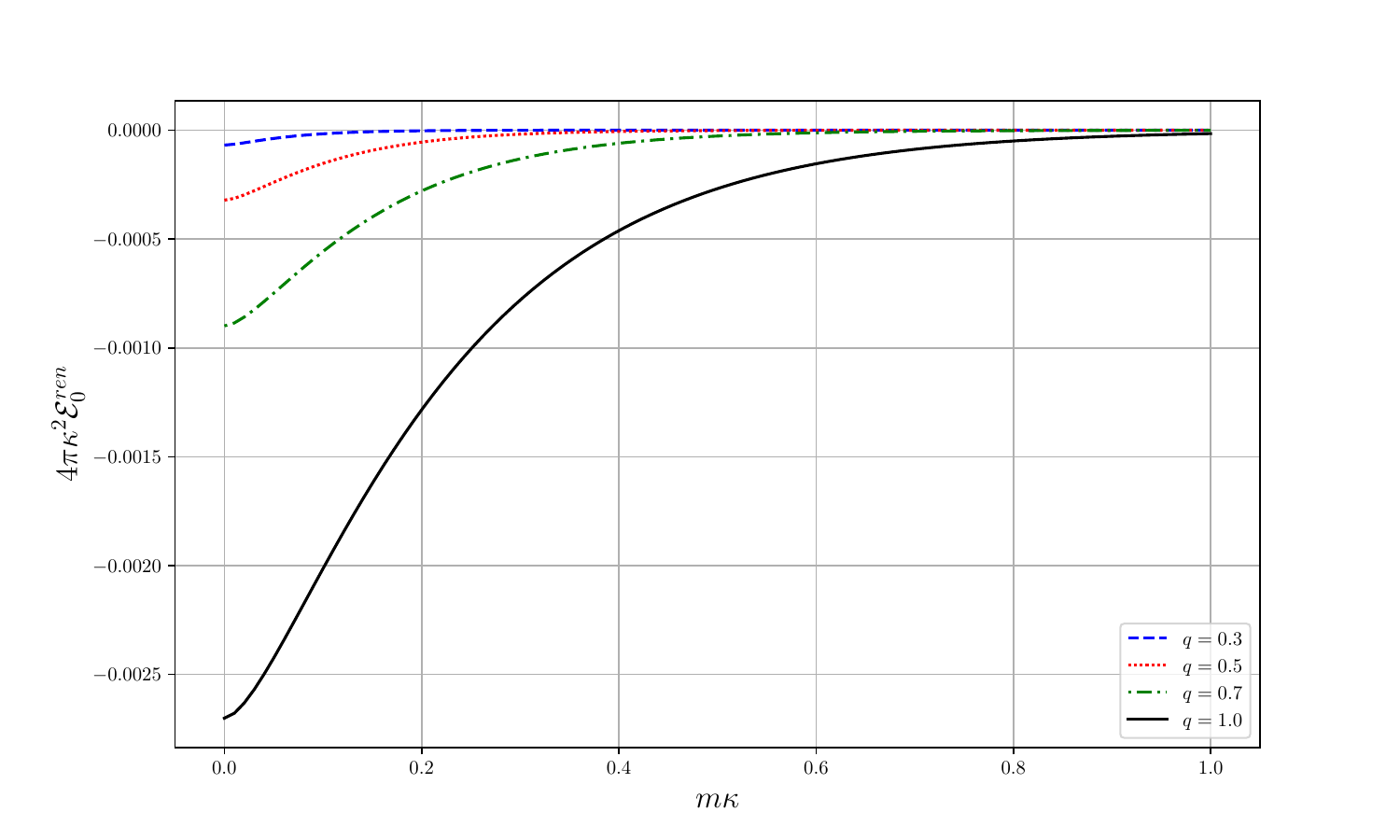}
     \includegraphics[scale=0.35]{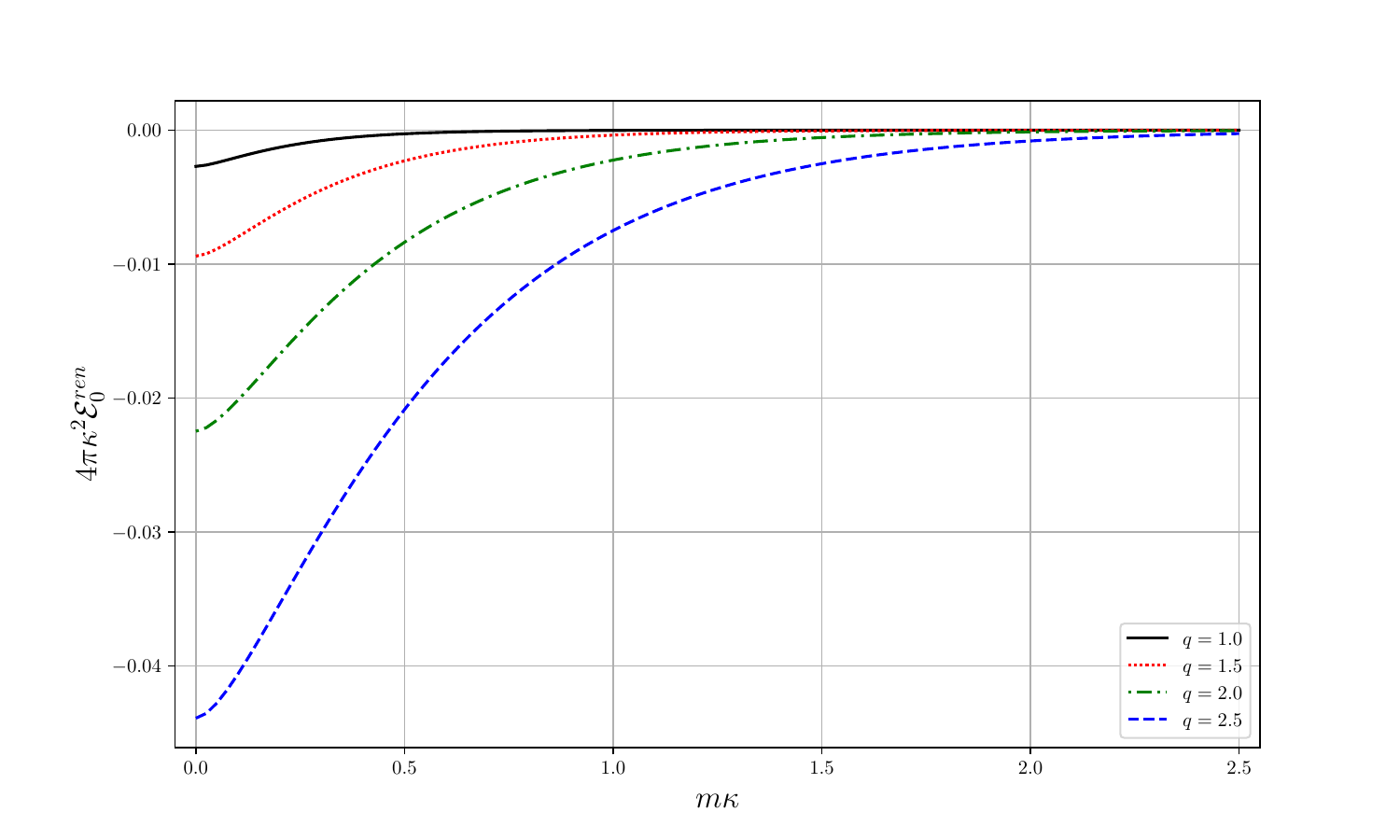}
    \caption{Plot of the vacuum energy density \eqref{renEdis} for $D=3$, considering different values of $q$.}
    \label{Czero1}
\end{figure}

The renormalized vacuum energy per unit `volume' $V_{D-2}$, from Eqs. \eqref{renEdis11}, \eqref{regularized}, \eqref{ren11} and \eqref{zetaReg}, is given by
\begin{eqnarray}\label{renEdis}
\mathcal{E}^{\text{ren}}_0(D) =-\frac{2\pi}{q}\frac{2^{\frac{D-1}{2}}m^{D-1}}{(4\pi)^{\frac{D+1}{2}}}\Biggl\{\sum_{\ell=1}^{[q/2]}\frac{f_{\frac{D-1}{2}}(m\bar{\kappa}\ell)}{s_{\ell}^2} - \frac{q}{2\pi^2}\int_0^\infty  \; \frac{dy}{s_y^2}\sideset{}{'}\sum_{k=-\infty}^{\infty}f_{\frac{D-1}{2}}(|m\bar{\kappa}k|)M_{k,q}(0, y)\Biggl\}.
 \end{eqnarray}
It is clear that the above expression satisfies the requirement \eqref{ren1} as a consequence of the asymptotic limit of the Macdonald function for large arguments, i.e., $K_{\mu}(z)\simeq\sqrt{\frac{\pi}{2z}}e^{-z}$ \cite{abramowitz, gradshteyn2000table}. Hence, in the limit $m\bar{\kappa}\rightarrow\infty$, the vacuum energy density above is exponentially suppressed. In Fig.\ref{Czero1} we have plotted, in terms of $m\kappa$, the vacuum energy density above for $D=3$. We can see that it is always negative and goes to zero as $m\kappa$ increases, in agreement with the condition \eqref{ren1}. In contrast, when $m\kappa\rightarrow 0$, the vacuum energy density \eqref{renEdis} converges to values associated with the massless case in Eq. \eqref{renEdis1} also for $D=3$. Note that for $q< 1$ the vacuum energy density increases and for $q> 1$ it decreases. The case $q=1$ is the vacuum energy density induced solely by the screw dislocation. 

The massless case of the vacuum energy density can be obtained by making use of the expression in Eq. \eqref{limite}. This provides
\begin{eqnarray}\label{renEdis1}
\mathcal{E}^{\text{ren}}_0(D) =-\left(\frac{q}{\pi}\right)^{D-2}\frac{\Gamma\left(\frac{D-1}{2}\right)}{(4\pi)^{\frac{D+1}{2}}}\frac{1}{\kappa^{D-1}}\Biggl\{\sum_{\ell=1}^{[q/2]}\frac{1}{\ell^{D-1}s_{\ell}^2} - \frac{q}{2\pi^2}\int_0^\infty  \; \frac{dy}{s_y^2}g(D,y,q)\Biggl\},
 \end{eqnarray}
 where
 \begin{eqnarray}\label{fun}
g(D,y,q) = \sideset{}{'}\sum_{k=-\infty}^{\infty}\frac{1}{k^{D-1}}M_{k,q}(0,y).
 \end{eqnarray}

 Note that the vacuum energy densities above are defined for $\kappa\neq0$. The case $\kappa=0$ needs to be considered from Eq. \eqref{heatglobal} and reproduces the conical spacetime discussion considered in the previous section. Also, in the massless scalar field case, $\bar{C}_{\frac{N}{2}}=0$ and we do not have the ambiguity present in the free energy defined in Eq. \eqref{CasimirEnDenZetaMethod}, neither it is needed a condition of the type \eqref{ren1}. Moreover, it is evident that the vacuum energy density in Eq. \eqref{renEdis1} vanishes in the limit $\kappa\rightarrow\infty$. In contrast, for $q=1$, we obtain only the screw dislocation contribution given by the second term in the r.h.s. of Eqs. \eqref{renEdis} and \eqref{renEdis1}.

 \subsection{Temperature corrections}
 We turn our analysis now to the temperature corrections to the vacuum energy densities \eqref{renEdis} and \eqref{renEdis1}. This is achieved by making use of Eqs. \eqref{TemExp} and \eqref{heatglobal2}. In the latter, the first two terms in the r.h.s. give contributions that depend on the heat kernel coefficients $C_0$ and $C_1$ and must be subtracted. For the massless scalar field case, these same two terms provide contributions of the type presented in Eq. \eqref{Temp_corr2} and, in particular, for $D=3$, of the type presented in Eq. \eqref{Temp_corr3} in which the term containing $C_0$ is the scalar black body radiation contribution \cite{Bezerra:2011nc, PhysRevD.83.104042, Mota:2022qpf}. 
  \begin{figure}[h]
    \includegraphics[scale=0.35]{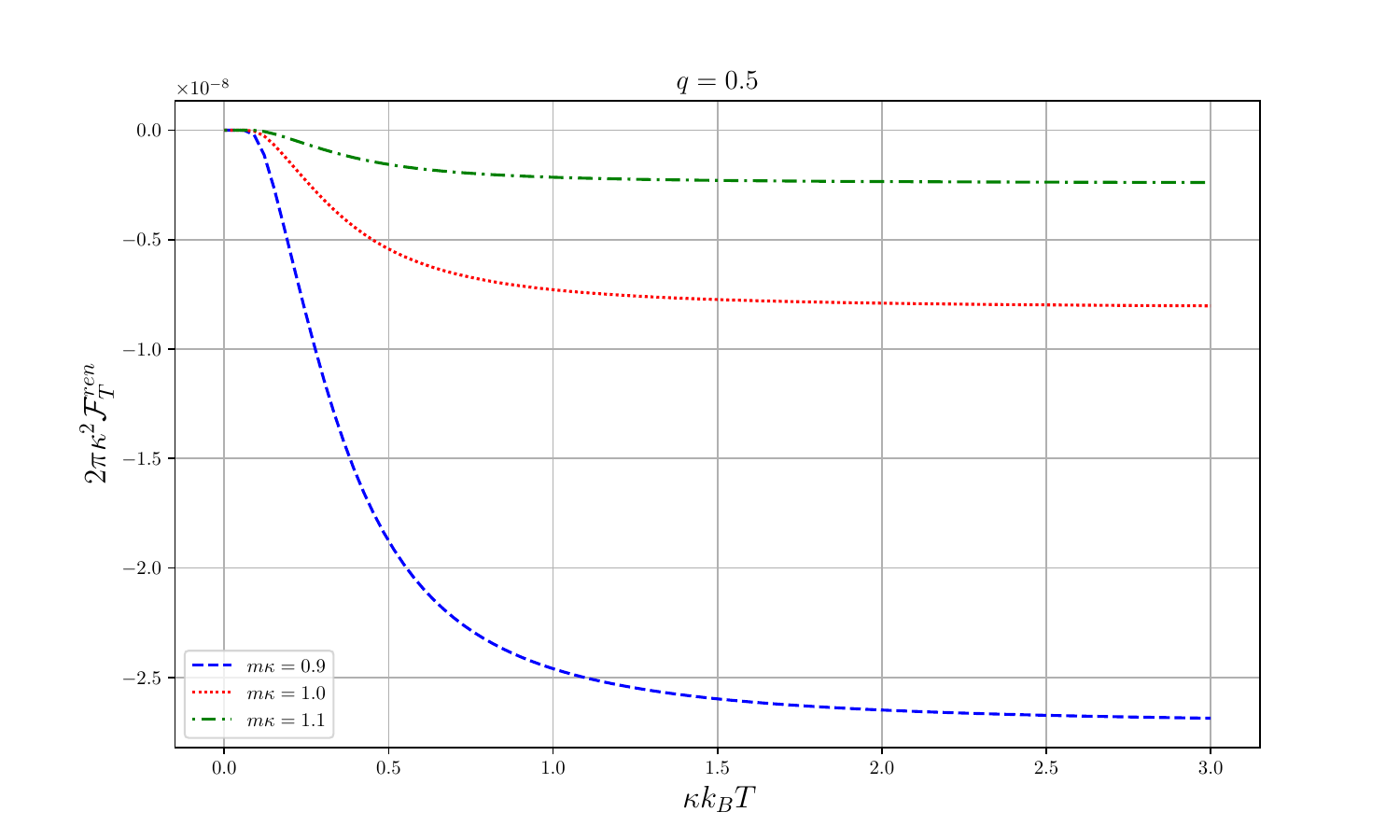}
     \includegraphics[scale=0.35]{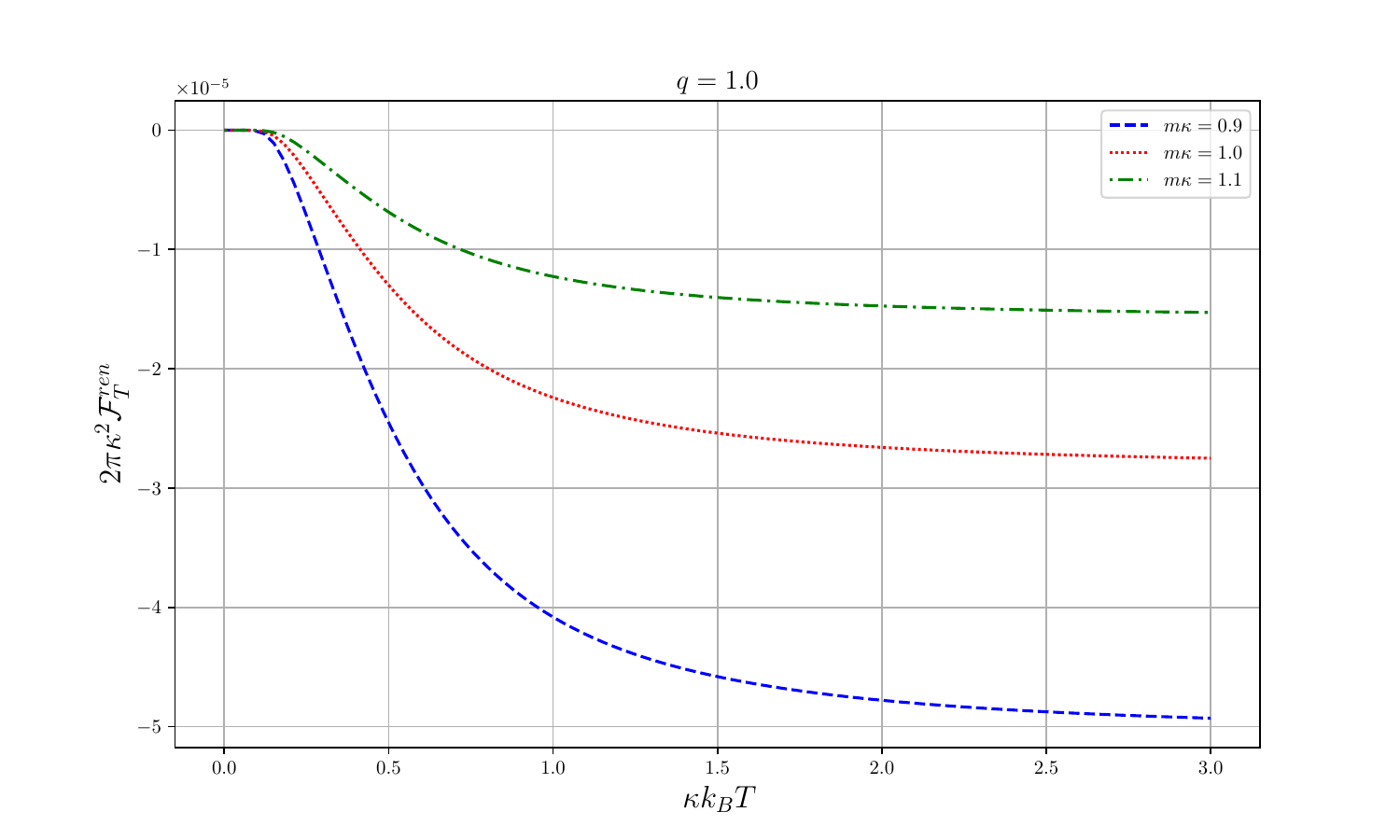}
      \includegraphics[scale=0.35]{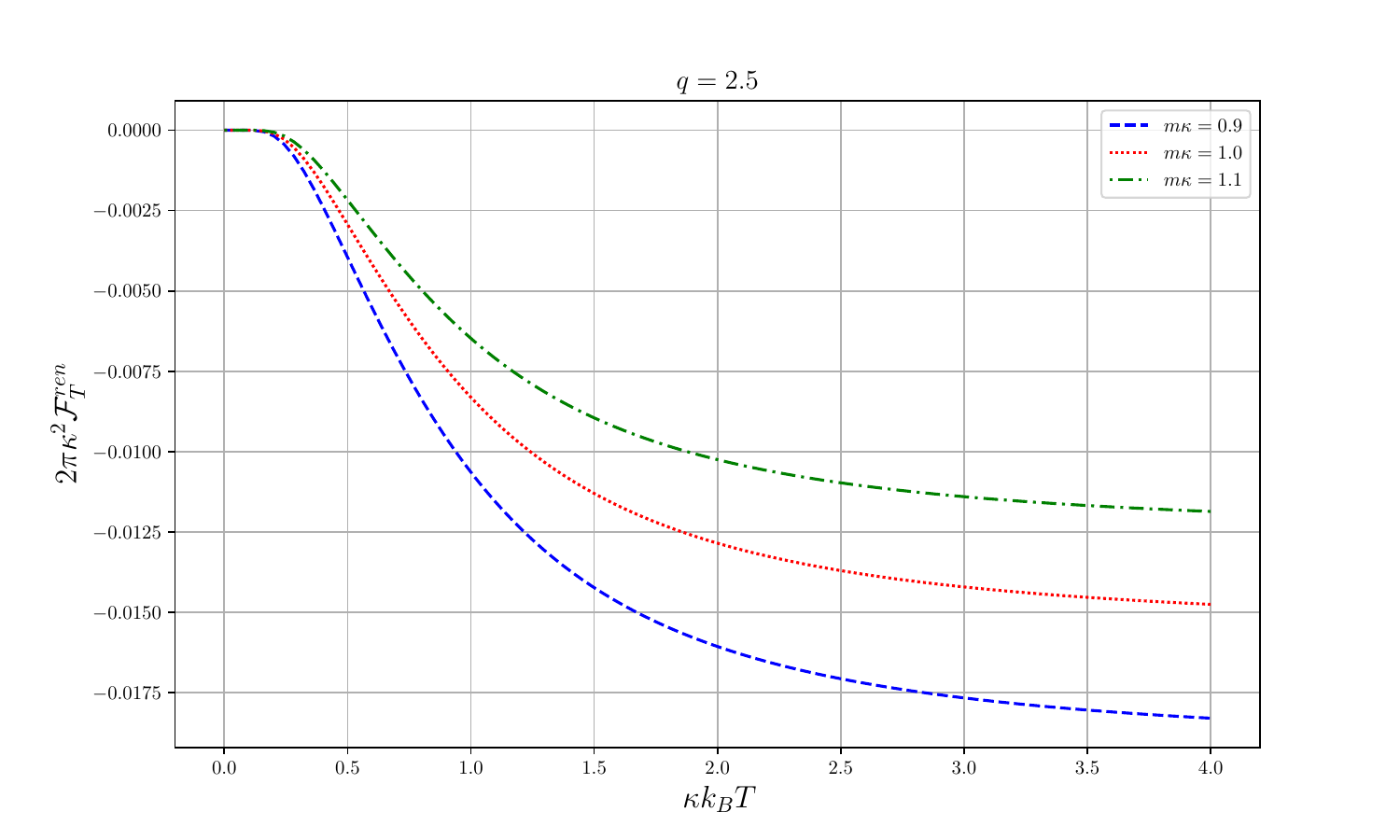}
       \includegraphics[scale=0.35]{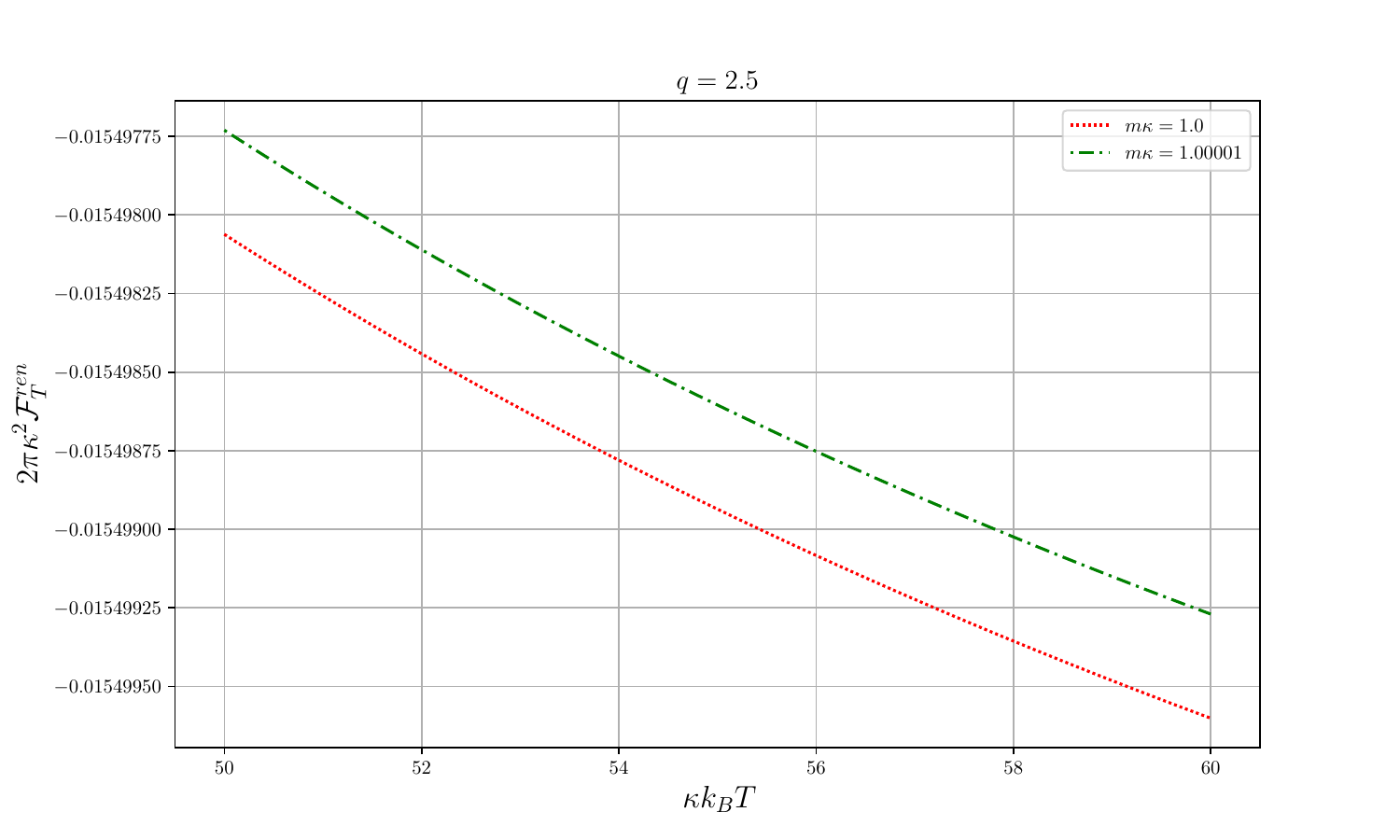}
    \caption{For $D=3$, plot of the free energy \eqref{renFECD} per unit of $V_1$, in terms of $\kappa k_BT$, for different values of $m\kappa$ and fixing four values of $q$.}
    \label{Czero}
\end{figure}
 
 By subtracting the terms discussed above, we have the renormalized free energy in the massive case as being 
\begin{eqnarray}\label{renFECD}
F^{\text{ren}}_T =-\frac{2^{\frac{D+3}{2}}\pi m^{D-1}}{q(4\pi)^{\frac{D+1}{2}}}V_{D-2}\sum_{n=1}^{\infty}\Biggl\{\sum_{\ell=1}^{[q/2]}\frac{f_{\frac{D-1}{2}}\left(mR_{n,\ell}\right)}{s_{\ell}^2} - \frac{q}{2\pi^2}\int_0^\infty  \; \frac{dy}{s_y^2} \sideset{}{'}\sum_{k=-\infty}^{\infty}f_{\frac{D-1}{2}}\left(mR_{n,k}\right)M_{k,q}(0, y)\Biggl\},
 \end{eqnarray}
where 
 \begin{eqnarray}\label{lengthscale}
R_{n,b} = \sqrt{(n\beta)^2 + (b\bar{\kappa})^2},
 \end{eqnarray}
with $b$ standing for $\ell$ in the first tem in the r.h.s. of \eqref{renFECD} and for $k$ in the second term.

Upon taking the limit of small arguments for the Macdonald function we can obtain, from Eq. \eqref{renFECD}, the massless case. This is given by
\begin{eqnarray}\label{renFECDMassless}
F^{\text{ren}}_T =-\frac{2^{D}\pi\Gamma\left(\frac{D-1}{2}\right)}{q(4\pi)^{\frac{D+1}{2}}}V_{D-2}\sum_{n=1}^{\infty}\Biggl\{\sum_{\ell=1}^{[q/2]}\frac{1}{s_{\ell}^2R^{D-1}_{n,\ell}} - \frac{q}{2\pi^2}\int_0^\infty  \; \frac{dy}{s_y^2}\sideset{}{'}\sum_{k=-\infty}^{\infty}\frac{M_{k,q}(0, y)}{R^{D-1}_{n,k}}\Biggl\}.
 \end{eqnarray}

 In particular, for $D=3$, we have 
 \begin{eqnarray}\label{renFECDMasslessD3}
F^{\text{ren}}_T =-\frac{1}{2\pi q}V_{1}\sum_{n=1}^{\infty}\Biggl\{\sum_{\ell=1}^{[q/2]}\frac{1}{s_{\ell}^2R^{2}_{n,\ell}} - \frac{q}{2\pi^2}\int_0^\infty  \; \frac{dy}{s_y^2}\sideset{}{'}\sum_{k=-\infty}^{\infty}\frac{M_{k,q}(0, y)}{R^{2}_{n,k}}\Biggl\},
 \end{eqnarray}
 where $V_1$ is in fact an infinite length in the $z$-direction. The expressions above for the free energy is completely convergent and depends on the conical parameter $q$, the screw dislocation parameter $\kappa$, the temperature $T$ and of the spatial dimension $D$. In Fig.\ref{Czero} we have plotted, for $D=3$, the free energy \eqref{renFECD} per unit of $V_1$ in terms of $\kappa k_BT$, assuming different values of $m\kappa$ and fixing four values of $q$. In each case, the plots show that the free energy goes to zero as $T\rightarrow 0$, as we should expect. In contrast, in the high-temperature limit, the free energy reaches the classical limit, that is, $F^{\text{ren}}_T \varpropto k_BT$, which is shown by the straight lines in the plot for $q=2.5$. The plots also show that the free energy increases as $m\kappa$ increases. 
 
 Below we shall analyze the interesting cases of the high- and low-temperature limits of the massless free energy for $D=3$, given by Eq. \eqref{renFECDMasslessD3}.
 
\subsubsection{High-temperature limit}
Let us first consider the high-temperature limit, $\pi\bar{\kappa}k_BT\gg 1$, of the expression in Eq. \eqref{renFECDMasslessD3}. For this, we shall write the sum in $n$ as 
  \begin{eqnarray}\label{identity1}
\frac{1}{\beta^2}\sum_{n=1}^{\infty}\frac{1}{\left[n^2 + \left(\frac{b\bar{\kappa}}{\beta}\right)^2\right]}
= -\frac{1}{2b^2\bar{\kappa}^2} + \frac{\pi k_BT}{2b\bar{\kappa}}\coth(b\bar{\kappa}\pi k_BT),
 \end{eqnarray}
 where again $b=\ell$ for the first term in the r.h.s. of the free energy \eqref{renFECDMasslessD3} while $b=k$ for the second term. Then, after substituting \eqref{identity1} in \eqref{renFECDMasslessD3}, for $\pi\bar{\kappa}k_BT\gg 1$, we have 
  \begin{eqnarray}\label{renFECDMasslessD3high}
F^{\text{ren}}_T \simeq-\frac{k_BT}{8\pi\kappa}V_1\Biggl\{\sum_{\ell=1}^{[q/2]}\frac{1}{\ell s_{\ell}^2} - \frac{q}{2\pi^2}\int_0^\infty  \; \frac{dy}{s_y^2}\sideset{}{'}\sum_{k=-\infty}^{\infty}\frac{M_{k,q}(0, y)}{k}\Biggl\}.
 \end{eqnarray}
A strightfoward dimensional analysis shows that the expression above is the classical limit of Eq. \eqref{renFECDMasslessD3}, as it should be. This corroborate with the fact that terms of quantum nature of the type \eqref{Temp_corr3} must be subtracted in order to obtain a renormalized free energy that has a dominating classical contribution at high temperatures. This asymptotic limit is confirmed by the straight lines in the plot of Fig.\ref{Czero} for $q=2.5$. In fact, this happens for any value of $q$.

\subsubsection{Low-temperature limit}
We want now to obtain an expression for low temperatures by considering the limit $\pi\bar{\kappa}k_BT\ll 1$ of Eq. \eqref{renFECDMasslessD3}. For this, it is convenient to re-write the sum in $n$ for the fist term in the r.h.s. of \eqref{renFECDMasslessD3} as follows 
   \begin{eqnarray}\label{identity2}
\sum_{n=1}^{\infty}\frac{1}{n^2\beta^2}\frac{1}{\left[1 + \left(\frac{\ell\bar{\kappa}}{n\beta}\right)^2\right]}= \frac{1}{\beta^2}\sum_{p=0}^{\infty}(-1)^p\left(\frac{\ell\bar{\kappa}}{\beta}\right)^{2p}\zeta_{\text{R}}(2p+2),
 \end{eqnarray}
 where a binomial expansion has been adopted for $\frac{\ell\bar{\kappa}}{n\beta}\ll 1$.
 
 On the other hand, for the second term in the r.h.s. of Eq. \eqref{renFECDMasslessD3} we can consider
    \begin{eqnarray}\label{identity3}
\frac{1}{\bar{\kappa}^2}\sum_{n=1}^{\infty}\sideset{}{'}\sum_{k=-\infty}^{\infty}\frac{M_{k,q}(0, y)}{\left[k^2 + \left(\frac{n\beta}{\bar{\kappa}}\right)^2\right]} \simeq \left[-\frac{2\pi^2}{q(\pi^2 + y^2)} - \frac{\pi\sin(\pi q)}{\cos(\pi q) - \cosh(qy)}\right](k_BT)^2\zeta_{\text{R}}(2) + \pi^2\kappa (k_BT)^3\zeta_{\text{R}}(3),
 \end{eqnarray}
 where we have first performed the sum in $k$. In the resulting expression we have taken a series expansion for $\frac{n\beta}{\bar{\kappa}}\gg 1$ and in the end performed the sum in $n$ for the first two leading terms of the expansion, which gave rise to the Riemann zeta functions in \eqref{identity3}.

Consequently, upon substituting Eqs. \eqref{identity2} and \eqref{identity3} in Eq. \eqref{renFECDMasslessD3}, up to third order in $k_BT$, we have 
 
   \begin{eqnarray}\label{renFECDMasslessDlow}
F^{\text{ren}}_T \simeq -\left[h(q,0) + \frac{\pi}{3q}\right]\frac{\zeta_{\text{R}}(2)}{4\pi^2}V_1(k_BT)^2 + \kappa\frac{\zeta_{\text{R}}(3)}{2\pi}V_1(k_BT)^3,
 \end{eqnarray}
 where $h(q,0)$ is a dimensionless function defined in Eq. \eqref{pole_ZF}. It is clear that in the absence of the conical defect, i.e., $q=1$, there is still contributions due to the screw dislocation, with $h(1,0)=0$. Again, a straightforward dimensional analysis shows that the terms present in Eq. \eqref{renFECDMasslessDlow} are of quantum nature and go to zero as $T\rightarrow 0$, which is clear in the plots of Fig.\ref{Czero} . Hence, in this limit, only the vacuum energy density at zero temperature \eqref{renEdis1} survives.

 In the system configuration studied in the present section, the heat kernel two-point function \eqref{Hkernel2C} in the coincidence limit retains information about the existing divergences, again, brought upon by the Minkowski spacetime flat nature, and also a divergence brought upon by the topological conical structure of the cosmic dispiration spacetime. The associated heat kernel coefficients are, respectively, given by Eq. \eqref{C0} and \eqref{HKCD1} and which are present in Eq. \eqref{heatglobal2}. Hence, after subtracting the terms containing these mentioned heat kernel coefficients by means of the renormalized process adopted here we obtain the vacuum energy and the corresponding temperature corrections. These results, as we have said before, present meaningful physical interpretation since they are not plagued with divergencies. Differently from the previous section where we consider a quasiperiodically identified conical spacetime, we obtain in the present section a nonzero result for the vacuum energy and temperature corrections, both in accordance with the normalization requirement given by Eq. \eqref{ren1}.

\section{Conclusion and discussion}
\label{sec5}
We have investigated scalar quantum vacuum fluctuations effects on the vacuum energy, temperature corrections and heat kernel coefficients arising from the nontrivial topology of $(D+1)$-dimensional quasiperiodically identified conical and cosmic dispiration spacetimes. In the quasiperiodically identified conical spacetime we have found a zero renormalized vacuum energy, as well as zero temperature corrections. The heat kernel coefficients obtained are the ones associated with the Euclidean divergence and also with the nontrivial topology of the spacetime. For the latter, as it is shown in Fig.\ref{HKC1}, we have learned that for some values of the quasiperiodic parameter $\alpha$ we obtain a zero value for the coefficient $C_1$, leaving only the coefficient $C_0$ related to the Euclidean divergence. Thus, we have generalized the results obtained in Ref. \cite{cognola:1993qg} for the heat kernel coefficient $C_1$ and recovered the author's results when $\alpha=0$, as it has been shown in Eq. \eqref{pole_ZF_q}. Also, the results that we have presented in Eqs. \eqref{RenVE} and \eqref{RenFE} are due to the renormalization scheme discussed in Refs. \cite{bordag2009advances, Bordag:1998rf}. 

Regarding the cosmic dispiration spacetime we have obtained, also adopting the renormalization scheme discussed in Refs. \cite{bordag2009advances, Bordag:1998rf}, a nonzero vacuum energy density, its corresponding temperature corrections and the heat kernel coefficients $C_0$, due to the Euclidean divergence, and $C_1$ due to the nontrivial topology divergence given by Eq. \eqref{HKCD1}. The induced vacuum energy density has been plotted in Fig.\ref{Czero1}, which shows that the renormalized vacuum energy density increases for $q<1$ and decreases for $q>1$. For $D=3$, in the massless scalar field case, we have presented expressions for the low and high temperature regimes. In the latter, we have shown that the free energy provides a pure classical contribution given by Eq. \eqref{renFECDMasslessD3high}, also confirmed by the straight lines in the plot of Fig.\ref{Czero} for $q=2.5$. The plots of Fig.\ref{Czero} also show that the temperature corrections go to zero as $T\rightarrow 0$, in agreement with the low-temperature expression in Eq. \eqref{renFECDMasslessDlow}.

The end this section, let us discuss some aspects and implications of our results. First of all, it is important to stress that spacetimes with conical singularities such as the spacetime of a cosmic string (disclination) and the spacetime of a cosmic dispiration considered here have been highly explored in the past decades. One of the reasons for the interest in these topological defects, besides the cosmological, astrophysical and gravitational consequences, described in the introduction, is that the corresponding spacetimes provide a curved background where it is possible in general to obtain nice results and analytical expressions. So, from the fundamental point of view this is always desirable, specially when one is dealing with curved backgrounds. The associated scalar two-point functions in these curved backgrounds as the ones obtained here, Eqs. \eqref{HeatKernel2pts} and \eqref{Hkernel2}, for instance, make possible to study phenomena such as entanglement harvesting \cite{Ji:2024fcq} and entanglement behavior of two static atoms \cite{Huang:2020rrj, Huang2020QuantumEI}. The two-point functions obtained here have also been used to investigate the vacuum expectation value of the energy-momentum tensor \cite{klecioQuasiPNanotubes, Braganca:2019mvj} and induced current density of a gauge field \cite{klecioQuasiPNanotubes, Braganca:2020jci, Braganca:2014qma}. 
 
 The implication of our results resides in the generalization of previous expressions found in literature, allowing us to better understand the fundamental structure of divergencies in the context of vacuum energy and its temperature corrections using the heat kernel coefficients method. A possible application of our results can be realized by using the two-point functions obtained to investigate entanglement phenomena.  Also, in a future work, we can use the vacuum energy and the temperature corrections found here to investigate backreaction effects in the energy-momentum tensor, similar to what have been done in Ref. \cite{DeLorenci:2008nr}.

{\acknowledgments}
The author is partially supported by the National Council for Scientific and Technological Development (CNPq) under grant No 311031/2020-0.

\end{document}